\documentclass[aps,prl,reprint]{revtex4-2}
\usepackage{bm}
\usepackage{graphicx}
\usepackage{subfigure}
\usepackage{amsmath}
\usepackage{amssymb}
\usepackage{color}
\usepackage[a4paper,left=1.5cm,right=1.3cm,top=3.0cm,bottom=2cm]{geometry}

\renewcommand{\Pr}   {\ifmmode \mathrm{Pr}  \else $\mathrm{Pr}$ \fi} 
\newcommand{\Ra}     {\ifmmode \mathrm{Ra}  \else $\mathrm{Ra}$ \fi} 
\newcommand{\Ro}     {\ifmmode \mathrm{Ro}  \else $\mathrm{Ro}$ \fi} 
\newcommand{\Ek}     {\ifmmode \mathrm{Ek}  \else $\mathrm{Ek}$ \fi} 
\newcommand{\Ta}     {\ifmmode \mathrm{Ta}  \else $\mathrm{Ta}$ \fi} 
\newcommand{\Nu}     {\ifmmode \mathrm{Nu}  \else $\mathrm{Nu}$ \fi} 
\renewcommand{\Re}   {\ifmmode \mathrm{Re}  \else $\mathrm{Re}$ \fi} 
\newcommand{\Fourier}{\ifmmode \mathcal{F}  \else $\mathcal{F}$ \fi} 
\newcommand{\real}   {\ifmmode \mathfrak{R} \else $\mathfrak{R}$ \fi} 
\newcommand{\imag}   {\ifmmode \mathfrak{I} \else $\mathfrak{I}$ \fi} 

\usepackage{comment}

\begin{document}
\title{Inverse cascades of kinetic energy and thermal variance in three-dimensional horizontally extended turbulent convection}
\author{Philipp P. Vieweg$^1$, Janet D. Scheel$^2$, Rodion Stepanov$^3$, and J\"org Schumacher$^{1,4}$}

\affiliation{$^1$ Institut f\"ur Thermo- und Fluiddynamik, Technische Universit\"at Ilmenau, D-98684 Ilmenau, Germany,\\
                $^2$ Department of Physics, Occidental College, 1600 Campus Road M21, Los Angeles, CA  90041, USA,\\
                $^3$ Institute of Continuous Media Mechanics, Korolyov 1, Perm, 614013, Russia,\\
                $^4$ Tandon School of Engineering, New York University, New York City, NY 11201, USA}
\date{\today}

\begin{abstract}
Inverse cascades of kinetic energy and thermal variance in the subset of vertically homogeneous modes in spectral space are found to cause a slow aggregation to a pair of convective supergranules that eventually fill the whole horizontally extended, three-dimensional, turbulent Rayleigh-B\'{e}nard convection layer when a heat flux is prescribed at the top and bottom. An additional weak rotation of the layer around the vertical axis stops this aggregation at a scale that is smaller than the lateral domain extension and ceases the inverse cascade for the thermal variance. The inverse cascade for the kinetic energy remains intact, even for times at which the root mean square values of temperature and velocity have reached the statistically steady state. This kinetic energy inverse cascade sustains the horizontally extended convection patterns which are best visible in the temperature field.  The resulting characteristic length of the aggregated convection patterns depends on the thermal driving and linearly on the strength of rotation. Our study demonstrates the importance of inverse energy cascades beyond the two-dimensional turbulence case in a three-dimensional convection flow that is subject to a multi-scale energy injection by thermal plumes and driven by boundary heat fluxes as typically present in natural geo- and astrophysical systems, such as solar convection.
\end{abstract}

\maketitle

\section{Introduction}
Turbulent convection processes in nature extend over many orders of magnitude in space and time and often exhibit hierarchies of cellular or vortical patterns such as in cloud clusters over tropical oceans \cite{Mapes1993}, storm systems on giant gas planets \cite{Young2017,Garcia2013}, or solar granulation and supergranulation \cite{Leighton1962,Dalsgaard2002,Rincon2018,Schumacher2020}. The transfer mechanisms of kinetic energy and thermal variance that sustain these hierarchical patterns are thus essential for a better understanding of their different length scales and lifetimes. The turbulent cascade model, which was suggested first by Richardson 100 years ago \cite{Richardson1922}, describes such a transfer of energy between eddies and plumes of different sizes, see also \cite{Gledzer1973,Frisch1978,Ohkitani1989,Eggers1991}. Depending on the dimensionality of the flow problem and the inclusion of further physical processes, such as rotation or magnetic fields, a cascade can be either {\em direct} or {\em inverse} \cite{Lesieur2008,Verma2018,Verma2019}. The inverse cascade that transfers energy from small to large scales was suggested first by Fj{\o}rtoft \cite{Fjortoft1953} and Kraichnan \cite{Kraichnan1967} for two-dimensional turbulence, where the fundamentally important vortex stretching mechanism is absent by the  dimensionality of the problem, thus leading to infinitely many further inviscid invariants, such as enstrophy or palinstrophy in addition to the turbulent kinetic energy \cite{Boffetta2012,Johnson2021}. Crossovers from an inverse to a direct cascade of kinetic energy were investigated numerically by adding strong rotation to three-dimensional homogeneous isotropic box turbulence \cite{Smith1996} or by a change of the size and dimensionality of the prescribed energy injection scale in three-dimensional thin-layer turbulence \cite{Celani2010,Poujol2020}. An existing inverse energy cascade may manifest in the formation of a big flow vortex condensate, which was observed in two-dimensional \cite{Smith1993,Frishman2017} or quasi-two-dimensional fluid turbulence \cite{Musacchio2019,vanKan2019}. Also, an inverse cascade can be initiated over scales where helicity is injected by external forces \cite{Plunian2020} in which case the turbulence remains three-dimensional, homogeneous and isotropic.

In turbulent thermal convection, the kinetic energy injection proceeds differently, namely in a whole range of scales rather than at a single prescribed scale \cite{Lohse2010}. This range incorporates the locally fluctuating thicknesses of the (detached) thermal boundary layer fragments -- the thermal plumes that drive the three-dimensional fluid turbulence and temperature field mixing \cite{Scheel2014,Schumacher2015}. The strong impact of such a multiscale forcing on a turbulent cascade is already known from a series of studies on fractal grid--generated fluid turbulence \cite{Seoud2007,Vassilicos2015}. 

\begin{figure*}[t]
\begin{center}
\includegraphics[scale=1.0]{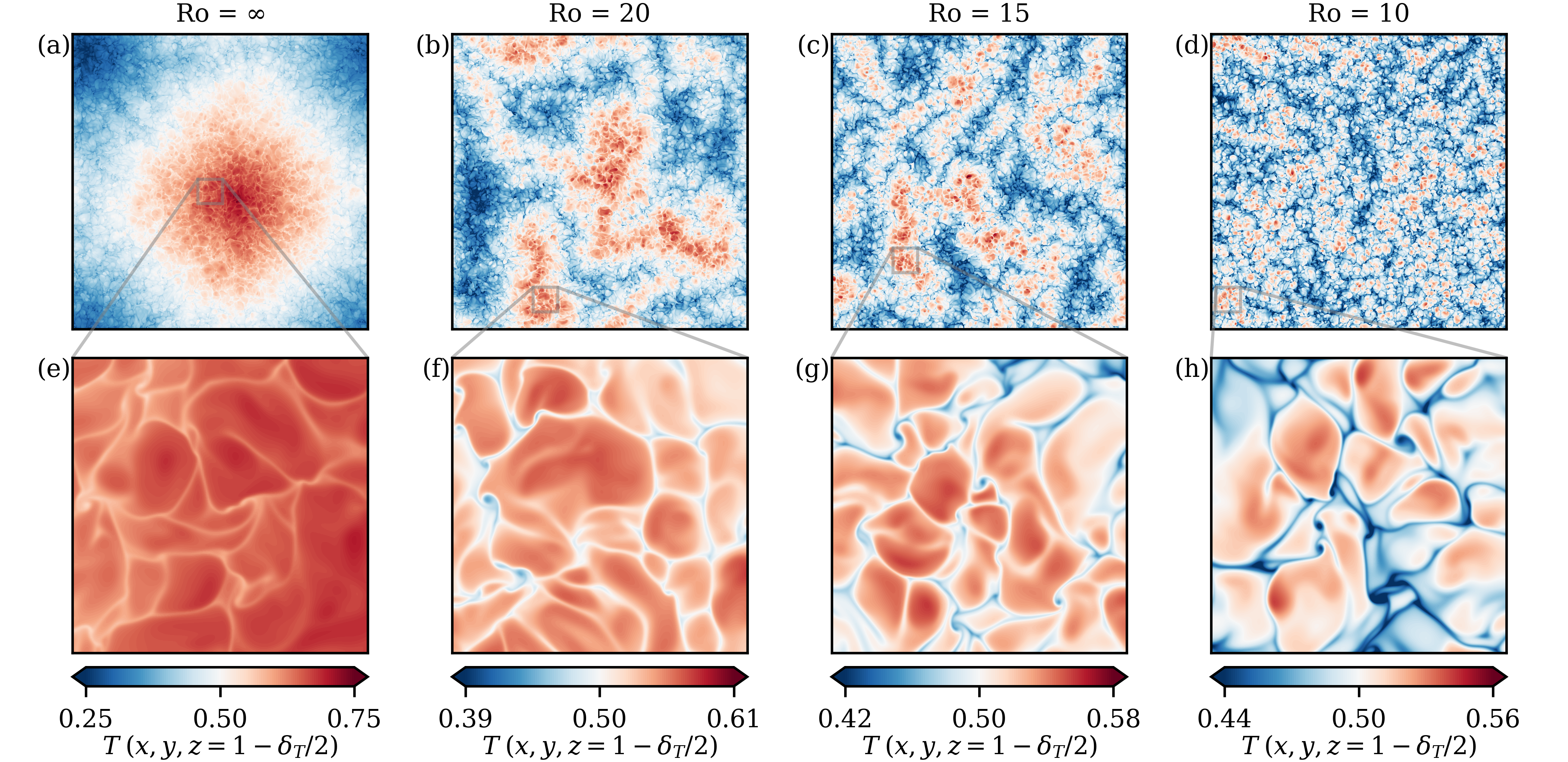}
\caption{Large-scale patterns of the temperature field for non-rotating and rotating scenarios. A view from the top close to the top surface inside the thermal boundary layer with thickness $\delta_{T} = 1 / ( 2 \Nu )$ is displayed. (a)--(d) Whole horizontal simulation cross-section of $60H \times 60H$ for four different Rossby numbers  $\Ro$ as given at the top of each panel. Rotation is -- starting from the non-rotating case on the left -- increased  to the right. (e)--(h) Magnifications of small regions of size $5H \times 5H$ that highlight the granules which are superposed to the supergranules and have a typical diameter of about 2--3 $H$. This suggests that locally the flow structure does not differ from the standard case with prescribed constant temperatures at the plates. All snapshots are taken in the final statistically stationary regime of the dynamical evolution for cases of $\Ra = 3.9 \times 10^{6}$. Corresponding color bars are given below each column. The pair of supergranules in panel (a) consists of the big hotter cell in the center of the plane and a colder cell that is distributed across the corners of the layer due to periodic boundary conditions with respect to $x$- and $y$-directions.}
\label{fig:1}
\end{center}
\end{figure*}

Convection cell formation in a three-dimensional Rayleigh-B\'{e}nard flow -- the fundamental setup of natural horizontally extended convection processes in a layer of height $H$ \cite{Kadanoff2001,Ahlers2009,Chilla2012} --  is found to be substantially determined by the choice of thermal boundary conditions, as shown recently in a series of direct numerical simulations (DNS) \cite{Vieweg2021}. For the case of constant heat flux boundary conditions at the top and bottom of the layer at $z=0$ and $z=H$, respectively, a slow aggregation to a pair of supergranular cells or supergranules was observed which is superposed upon a network of smaller granular convection cells of an extension $\mathcal{O} (H)$ for different velocity boundary conditions. Both the linear stability analysis and weakly nonlinear theory right above the onset of thermal convection predict the formation of a such a pair of counter-rotating convection cells that fill the entire domain \cite{Sparrow1964,Hurle1967,Busse1967,Chapman1980}. Interestingly,  a leading Lyapunov vector analysis, which quantifies the sensitivity of the convection flow trajectory in the high-dimensional phase or state space with respect to small perturbations, demonstrated clearly that this behavior continues to exist in the fully turbulent regime \cite{Vieweg2021}. The system has thus not forgotten about its primary instability mechanism; instabilities at a certain scale $\ell$ give rise to structures at new scales $\ell^{\prime}>\ell$ until domain size $L$ is reached. Since this growth proceeds for all accessible Rayleigh numbers (which quantify the strength of thermal driving of the flow) ${\rm Ra}\lesssim 10^8$, it is natural to ask  which cascade processes this growth is based on and which additional physical mechanisms can stop this (unnatural) growth to full domain size. 

In the present work, we trace this supergranule aggregation in a non-rotating, horizontally extended, three-dimensional, turbulent Rayleigh-B\'{e}nard flow -- which is driven by a constant heat flux at the top and bottom -- back to inverse cascades in the Fourier spectral space for both the turbulent kinetic energy and thermal variance. It is demonstrated that both cascades are established dominantly in the subset of two-dimensional Fourier modes, i.e., with  wave vectors obeying a vertical component $k_z=0$. These inverse cascades are found to cover an extended range of horizontal wave numbers. We show that the inverse cascade of the thermal variance exists as long as the supergranule continues to grow. It ceases once it occupies the entire domain. Conversely, the inverse cascade of the kinetic energy proceeds to exist and sustains the pair of supergranules in the domain. Moreover, we show that an additional {\it weak rotation} with a rotation rate vector ${\bm \Omega}=\Omega {\bm e}_z$ limits the gradual aggregation at an intermediate scale $\Lambda$ such that $H < \Lambda < L$ with $L$ and $H$ being the horizontal and vertical extension of the convection layer, respectively. We show that the resulting large-scale pattern size depends linearly on the strength of rotation. This behavior is underlined by Figs. \ref{fig:1} (a)--(d) where the differently strong filamentation of the supergranule is demonstrated for different dimensionless Rossby numbers $\Ro$ which measure the weakness of rotation. The magnifications in Figs. \ref{fig:1} (e)--(h) also show that the superposed granules -- the convection cells with an extension of roughly $\Lambda \approx 3$ -- continue to exist during this significant change of the large-scale organisation of convection. This suggests that the local mechanisms in the boundary layer, such as the plume detachment, remain qualitatively similar to those in the standard Rayleigh-B\'{e}nard case with prescribed temperatures at the top and bottom.

Our study thus explores in detail the formation of hierarchical convection patterns in a simple paradigm for atmospheric and astrophysical mesoscale convection where typically heat fluxes rather than prescribed temperatures drive the turbulence. Moreover, it sheds a new light on the relevance and existence of inverse cascades in fully three-dimensional natural flows that typically receive their energy by a whole range of scales. Furthermore, the present study demonstrates how quasi-two-dimensional structures in three-dimensional turbulence can be formed, and how their size is controlled by \textit{weak} rotation as an additional physical process. Our work thus has interesting implications for a better fundamental understanding of atmospheric turbulence or solar convection \cite{Schumacher2020,Vasil2021}.

\begin{figure}[t]
\begin{center}
\includegraphics[scale=1.0]{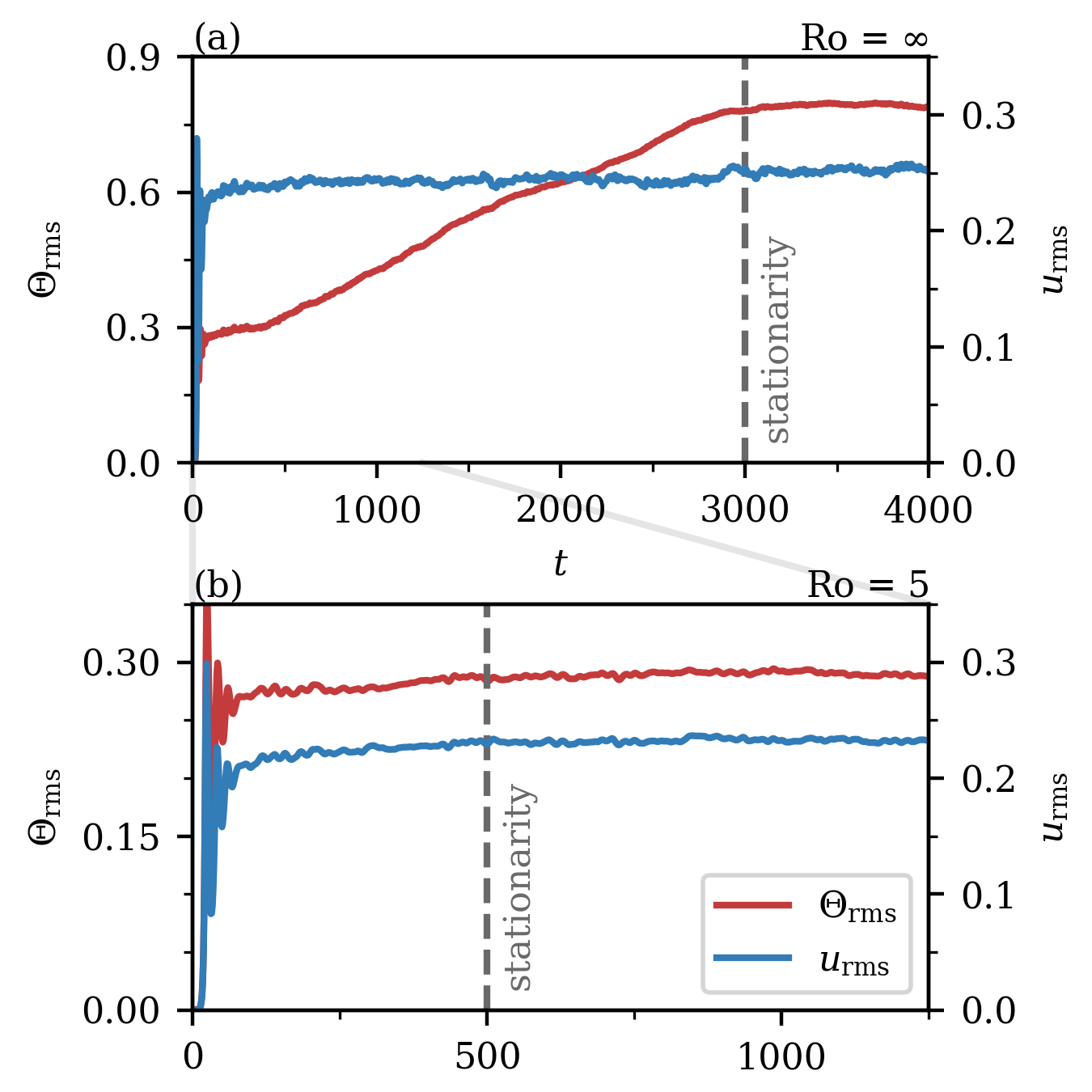}
\caption{Evolution of the root-mean-square (rms) values of the temperature and velocity fields. (a) Non-rotating run R1. (b) Weakly rotating run R1\_Ro5 at $\Ro = 5$. The vertical dashed lines indicate the times at which the simulations reach a statistically quasi-steady regime. All data are for $\Ra = 1.0 \times 10^{4}$, see table \ref{tab:1}.}
\label{fig:2}
\end{center}
\end{figure}

\section{Numerical simulations} 
The structure formation is investigated by comprehensive numerical simulations. We solve the equations of three-dimensional Rayleigh-Bénard convection in the Boussinesq approximation, including a Coriolis force term,  by applying the spectral element method Nek5000 \cite{Fischer1997,Scheel2013,Vieweg2021}. The equations are non-dimensionalized by the layer height $H$, the free-fall velocity $U_f=\sqrt{\alpha g \beta H^2}$, and a temperature $\beta H$ with the prescribed temperature gradient $\beta=-\partial T/\partial z$ at $z=0,H$ \cite{Johnston2009,Foroozani2021}. Here, $\alpha$ and $g$ are the thermal expansion coefficient and the acceleration due to gravity, respectively. Times are measured in characteristic free-fall time units $\tau_{f}=H/U_f$. For simplicity, we term $\tilde{\Theta}$ -- the deviation of the total temperature field from the linear conduction profile -- in the following as the temperature field. The governing equations are given by
\begin{align}
\label{eq:CE}
\tilde{\nabla} \cdot \tilde{\bm{u}} & = 0\,, \\
\label{eq:NSE}
\frac{\partial \tilde{\bm{u}}}{\partial \tilde{t}} + ( \tilde{\bm{u}} \cdot \tilde{\nabla} ) \tilde{\bm{u}} + \frac{1}{\Ro} \bm{e}_{z} \times \tilde{\bm{u}} & = - \tilde{\nabla} \tilde{p} + \sqrt{\frac{\Pr}{\Ra}} \tilde{\nabla}^{2} \tilde{\bm{u}} + \tilde{\Theta} \bm{e}_{z}\,, \\
\label{eq:EE}
\frac{\partial \tilde{\Theta}}{\partial \tilde{t}} + ( \tilde{\bm{u}} \cdot \tilde{\nabla} ) \tilde{\Theta} & = \frac{1}{\sqrt{\Ra \Pr}} \tilde{\nabla}^{2} \tilde{\Theta} + \tilde{u}_{z} \,.
\end{align}
Note that while the Coriolis force is included explicitly, the centrifugal force is absorbed in the pressure term \cite{Verma2018}. Three dimensionless parameters control the whole dynamics. These parameters are the Prandtl number, which characterizes the fluid itself, and the Rayleigh and Rossby numbers, all of which are given by
\begin{equation}
\Pr = \frac{\nu}{\kappa}=1 , \quad
\Ra = \frac{\alpha g \beta H^{4}}{\nu \kappa} , \quad
\Ro = \frac{U_{f}}{2 \Omega H} \gtrsim 2.5
\label{params}
\end{equation}
with the kinematic viscosity of the fluid $\nu$, the temperature diffusivity $\kappa$, and the rotation rate $\Omega$.  The Rayleigh number varies between $1.0 \times 10^{4} \leq \Ra \leq 7.7 \times 10^{7}$. The Cartesian simulation domain is $V = L \times L \times H$ with an aspect ratio $\Gamma = L/H = 60$ for all cases. From now on, we omit the tilde for the non-dimensional quantities for a simpler notation. Periodic boundary conditions are applied in the horizontal directions $x$ and $y$. At the bottom and top planes at $z=0,1$, free-slip velocity boundary conditions for ${\bm u}=(u_x, u_y, u_z)$
\begin{equation}
\label{eq:free_slip_BC}
\frac{\partial u_x}{\partial z}\Bigg|_{z=0,1} = \frac{\partial u_y}{\partial z}\Bigg|_{z=0,1}=0 \quad \text{and} \quad u_z\big|_{z=0,1}=0 \,
\end{equation}
and Neumann-type constant heat flux boundary conditions for the temperature field are taken,
\begin{equation}
\label{eq:thermal_BC_Theta}
\frac{\partial \Theta}{\partial z} \Bigg|_{z=0,1} = 0 \,.
\end{equation}
The DNS are performed on a non-uniformly spaced grid. For the subsequent spectral analysis, we thus interpolate the fields with spectral accuracy onto a uniform 3d mesh. We refer to ref. \cite{Vieweg2021} for more details on the simulations and list in Table \ref{tab:1} further parameters of the simulations, such as the Nusselt and Reynolds numbers, $\Nu = 1 + \sqrt{\Ra \Pr} \thickspace \langle u_{z} T \rangle$ and $\Re = \sqrt{\Ra / \Pr} \thickspace u_{\mathrm{rms}}$ where $\langle\cdot\rangle$ denotes a combined volume-time average, which quantify the turbulent heat and momentum transport, respectively. We also provide the total lengths of the time evolution of the different convection flows.

All simulations start from the diffusive equilibrium state which is randomly and infinitesimally perturbed in the temperature field. We plot in Fig. \ref{fig:2} the evolution of the root-mean-square (rms) temperature and velocity, $\Theta_{\text{rms}}(t) = \langle \Theta^2 ({\bm x},t) \rangle_V^{1/2}$ and $u_{\text{rms}}(t) = \langle \bm{u}^2 ({\bm x},t) \rangle_V^{1/2}$, respectively. Although both fields are connected via \eqref{eq:NSE} and \eqref{eq:EE}, their evolution differs significantly. On the one hand, the supergranule aggregation process is connected to a prominent long-term linear increase of the temperature fluctuations that proceeds over thousands of free-fall times in the non-rotating case, see the red curve in Fig. \ref{fig:2}(a). Once this aggregation comes to an end, the temperature field relaxes into a statistically steady regime. In contrast on the other hand, $u_{\rm rms}$ is already in a statistically steady regime after a much shorter initial period as seen by the blue curve in (a). In the rotating case, see Fig. \ref{fig:2}(b), the supergranule formation can be suppressed and thus the long-term increase of thermal variance shortens depending on $\Ro$.

We have also calculated the anisotropy parameter $A = E_{\mathrm{hor}} / (2 E_{\mathrm{vert}})$ \cite{Verma2017a} with $E_{\mathrm{hor}} = \langle (u_{x}^{2} + u_{y}^{2})/2 \rangle$ and $E_{\mathrm{vert}} = \langle u_{z}^{2}/2 \rangle$ for the statistically stationary state and find $A_{\mathrm{Nfs1}} = 4.8$, $A_{\mathrm{Nfs3}} = 1.9$, $A_{\mathrm{Nfs4}} = 1.7$, see Table \ref{tab:1}. The simulations which correspond to runs Nfs1 and Nfs3 with Dirichlet rather than Neumann boundary conditions for the temperature field give smaller anisotropy factors of $A_{\mathrm{Dfs1}} = 1.9$, $A_{\mathrm{Dfs3}} = 1.4$ \cite{Vieweg2021}. Hence, albeit the Neumann flow tends to offer a stronger anisotropy, both flows exhibit roughly similar anisotropies which underlines the coexistence of different hierarchical flow structures in the Neumann case. Furthermore, we observe that the anisotropy decreases with increasing Rayleigh number and thus an increased level of turbulence.

\begin{table*}[t]
\centering
\begin{tabular}{l r r r r r r r r r}
\hline\hline
\multicolumn{1}{l}{Run}	& \multicolumn{1}{c}{\Ra} & \multicolumn{1}{c}{\Ro}	& \multicolumn{1}{c}{$N_{e}$}	& \multicolumn{1}{c}{$N$}	& \multicolumn{1}{c}{$t_{\text{tot}} (\tau_{f})$}	& \multicolumn{1}{c}{$t_{\text{tot}} (\tau_{\nu})$}	& \multicolumn{1}{c}{${\rm Nu}$}	& \multicolumn{1}{c}{${\rm Re}$} 	& \multicolumn{1}{c}{$\Lambda$}\\
\hline
R1		    & $1.04 \times 10^{4}$	& $\infty$	& $160,000$		& $7$	& $ 4,000$	& $39.2$	& $3.96 \pm 0.10$	& $26.5 \pm 0.3$	& $59.7 \pm 0.0$\\
R1\_Ro5	    & $1.04 \times 10^{4}$	& $5$		& $160,000$		& $7$	& $ 1,250$	& $12.2$	& $3.69 \pm 0.04$	& $24.4 \pm 0.2$	& $21.9 \pm 3.3$\\
\\
R2		    & $2.04 \times 10^{5}$	& $\infty$	& $160,000$		& $11$	& $ 6,500$	& $14.4$	& $6.74 \pm 0.10$	& $81.4 \pm 0.7$	& $59.7 \pm 0.0$\\
R2\_Ro10	& $2.04 \times 10^{5}$	& $10$		& $160,000$		& $9$	& $ 4,500$	& 	        & $6.61 \pm 0.05$	& $80.3 \pm 0.8$	& $42.0 \pm 0.7$\\
R2\_Ro9	    & $2.04 \times 10^{5}$	& $9$		& $160,000$		& $9$	& $ 4,500$	& 	        & $6.57 \pm 0.06$	& $78.7 \pm 0.9$	& $33.2 \pm 1.8$\\
R2\_Ro8	    & $2.04 \times 10^{5}$	& $8$		& $160,000$		& $9$	& $ 4,500$	& 	        & $6.53 \pm 0.06$	& $76.3 \pm 0.5$	& $29.0 \pm 0.9$\\
R2\_Ro7	    & $2.04 \times 10^{5}$	& $7$		& $160,000$		& $9$	& $ 4,500$	& 	        & $6.45 \pm 0.05$	& $72.8 \pm 0.3$	& $25.0 \pm 0.7$\\
R2\_Ro6	    & $2.04 \times 10^{5}$	& $6$		& $160,000$		& $9$	& $ 4,500$	& 	        & $6.32 \pm 0.04$	& $69.1 \pm 0.3$	& $18.4 \pm 0.4$\\
R2\_Ro5	    & $2.04 \times 10^{5}$	& $5$		& $160,000$		& $9$	& $ 4,500$	& 	        & $6.13 \pm 0.03$	& $65.1 \pm 0.3$	& $16.2 \pm 1.4$\\
 \\
R3		    & $3.93 \times 10^{6}$ 	& $\infty$	& $1,280,000$	& $7$	& $10,000$	& $5.0$	    & $12.29 \pm 0.16$	& $229.0 \pm 1.4$	& $59.7 \pm 0.0$\\
R3\_Ro30	& $3.93 \times 10^{6}$ 	& $30$		& $1,280,000$	& $7$	& $ 4,500$	& 	        & $12.38 \pm 0.07$	& $285.9 \pm 1.3$	& $59.2 \pm 0.1$\\
R3\_Ro20	& $3.93 \times 10^{6}$ 	& $20$		& $1,280,000$	& $7$	& $ 4,500$	& 	        & $12.27 \pm 0.04$	& $231.5 \pm 1.0$	& $45.6 \pm 0.2$\\
R3\_Ro17	& $3.93 \times 10^{6}$ 	& $17$		& $1,280,000$	& $7$	& $ 4,500$	& 	        & $12.26 \pm 0.06$	& $231.4 \pm 1.6$	& $36.7 \pm 2.1$\\
R3\_Ro15	& $3.93 \times 10^{6}$ 	& $15$		& $1,280,000$	& $7$	& $ 4,500$	& 	        & $12.23 \pm 0.05$	& $224.5 \pm 1.3$	& $28.7 \pm 1.4$\\
R3\_Ro13	& $3.93 \times 10^{6}$ 	& $13$		& $1,280,000$	& $7$	& $ 4,500$	& 	        & $12.18 \pm 0.04$	& $217.4 \pm 1.2$	& $25.3 \pm 0.5$\\
R3\_Ro10	& $3.93 \times 10^{6}$ 	& $10$		& $1,280,000$	& $7$	& $ 4,500$	& 	        & $12.05 \pm 0.04$	& $205.1 \pm 0.9$	& $19.8 \pm 1.1$\\
R3\_Ro10s	& $3.93 \times 10^{6}$ 	& $10$		& $1,280,000$	& $7$	& $ 4,500$	& $2.3$     & $12.04 \pm 0.04$	& $199.7 \pm 0.8$	& $17.9 \pm 1.6$\\
R3\_Ro9	    & $3.93 \times 10^{6}$ 	& $9$		& $1,280,000$	& $7$	& $ 4,500$	& 	        & $11.97 \pm 0.04$	& $199.6 \pm 1.1$	& $15.4 \pm 0.4$\\
R3\_Ro8	    & $3.93 \times 10^{6}$ 	& $8$		& $1,280,000$	& $7$	& $ 4,500$	& 	        & $11.85 \pm 0.04$	& $197.5 \pm 0.7$	& $11.4 \pm 0.3$\\
R3\_Ro7	    & $3.93 \times 10^{6}$ 	& $7$		& $1,280,000$	& $7$	& $ 4,500$	& 	        & $11.72 \pm 0.03$	& $205.9 \pm 0.8$	& $10.4 \pm 0.4$\\
 \\
R4		    & $7.69 \times 10^{7}$	& $\infty$	& $11,022,400$	& $7$	& $19,000$	& $2.2$	    & $23.47 \pm 0.24$	& $635.9 \pm 3.1$	& $59.7 \pm 0.0$\\
R4\_Ro30	& $7.69 \times 10^{7}$	& $30$		& $11,022,400$	& $7$	& $ 7,500$	& 	        & $23.44 \pm 0.07$	& $640.3 \pm 3.3$	& $27.3 \pm 0.5$\\
\hline\hline
\end{tabular}
\caption{Parameters of all direct numerical simulation runs. We list the Rayleigh number $\Ra$, the Rossby number $\Ro$, the total number of spectral elements in the simulation domain $N_{e}$, the polynomial order $N$ on each spectral element and with respect to each space dimension, and the total dimensionless runtime of the simulation $t_{\text{tot}}$ in units of the corresponding free-fall times $\tau_{f}$. For those simulations which started with a linear equilibrium profile of the temperature and the fluid at rest, the total runtime is also translated into vertical viscous diffusion times $\tau_{\nu}=\sqrt{{\rm Ra}/{\rm Pr}} \,\tau_f$. For the other simulations, the final non-rotating field states are taken as initial conditions. For all simulations, the Prandtl number $\Pr = 1$. We also list the resulting Nusselt number $\Nu$, the Reynolds number $\Re$, as well as the integral length scale $\Lambda$ based on the vertically averaged temperature field, see eq. \eqref{intscale}. All values of these three statistical quantities are obtained from at least $50$ snapshots taken during the last 500 $\tau_{f}$. For runs R1 and R1\_Ro5, these statistics are obtained from even longer times intervals, see Fig. \ref{fig:2}. This is similar for run R3\_Ro10s which, in contrast to run R3\_Ro10, started from a linear equilibrium temperature profile. This run demonstrates that statistical results are independent of the initial conditions. Error bars are determined from the standard deviation.}
\label{tab:1}
\end{table*}

\section{Inverse cascade in non-rotating case} 
\begin{figure*}[ht!]
\begin{center}
\includegraphics[scale=1.0]{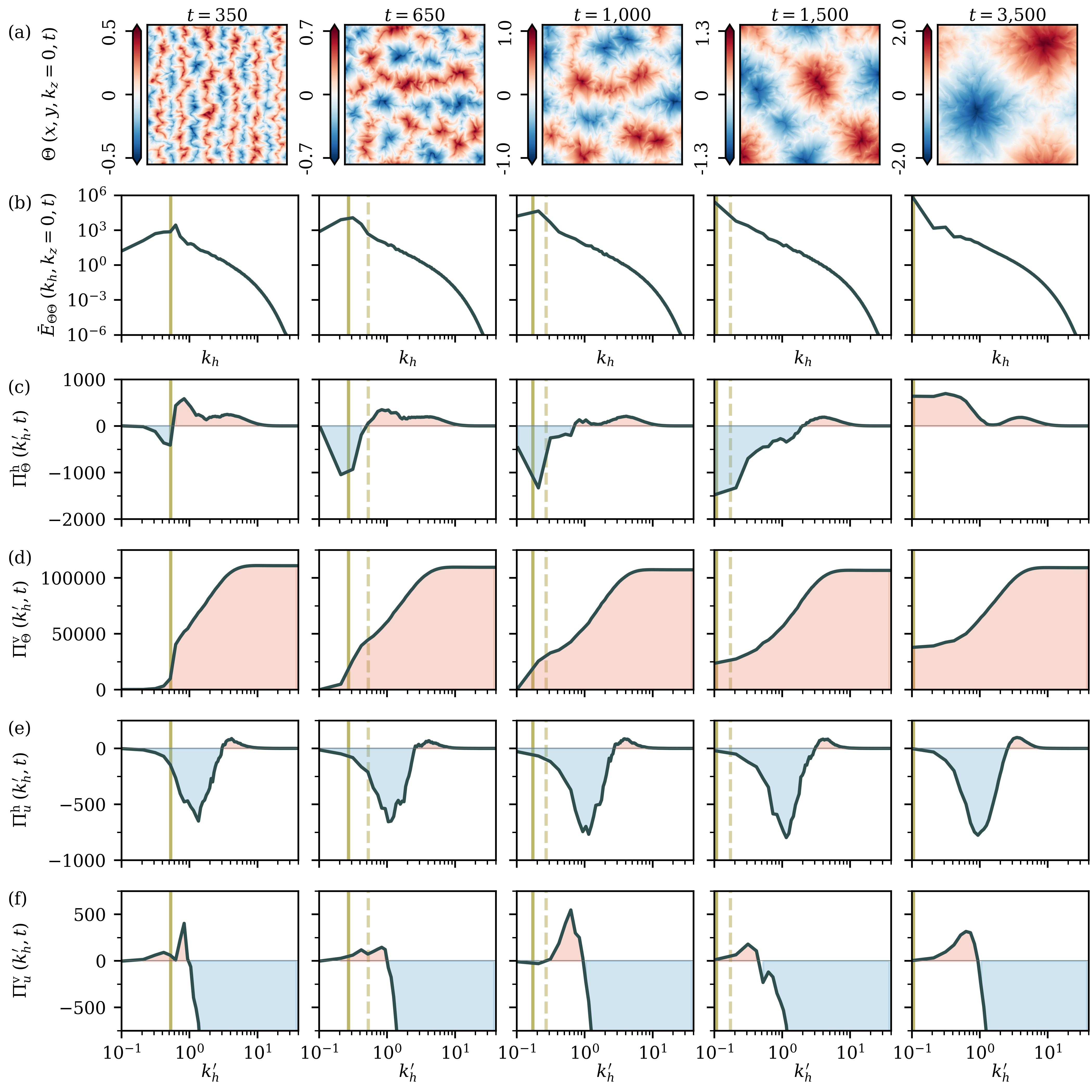}
\caption{Spectral transfer analysis for the long-term simulation R1 at a Rayleigh number $\Ra = 1.0 \times 10^{4}$ without rotation. 
(a) Instantaneous vertically averaged temperature field at different times. 
(b) Corresponding azimuthally averaged thermal variance spectrum. 
The solid vertical lines in these and all subsequent panels indicate the wave number that corresponds to the typical pattern size at the given time, see eq. \eqref{intscale}, whereas the dashed lines replot this quantity from the previous panel to highlight the growth of this length scale over time. 
(c) Planar spectral thermal variance flux $\Pi^{\mathrm{h}}_{\Theta} (k_{h}^{\prime}, t)$. 
(d) Three-dimensional spectral thermal variance flux $\Pi^{\mathrm{v}}_{\Theta} (k_{h}^{\prime}, t)$. 
(e) Planar spectral kinetic energy flux $\Pi^{\mathrm{h}}_u (k_{h}^{\prime}, t)$. 
(f) Three-dimensional spectral kinetic energy flux $\Pi^{\mathrm{v}}_{u} (k_{h}^{\prime}, t)$. 
The data in the last column are obtained when the supergranule growth ceased and the turbulent flow is statistically stationary -- the rightmost panels of (b)--(f) are thus time-averaged over $1000  \tau_{f}$ centered around the given time. In panels (b)--(f), blue (red) filled areas indicate an inverse (direct) cascade.}
\label{fig:3}
\end{center}
\end{figure*}

In the non-rotating convection case, the Rossby number is $\Ro=\infty$. The cascade process is analysed in detail in Fig. \ref{fig:3} for run R1 at a Rayleigh number $\Ra = 1.0 \times 10^{4}$ as an example. The supergranule aggregation is seen to proceed in the horizontal direction \cite{Vieweg2021}. Since its horizontal extension is much larger than the height of the layer, we visualize in the first row of panels, Figs. \ref{fig:3}(a), the vertically averaged temperature field which again demonstrates the prominent slow aggregation towards the supergranule. Using a spectral expansion of all fields (see Appendix A for the Neumann case or ref. \cite{Verma2019}), we can derive the modal evolution equations for kinetic energy and thermal variance. The time evolution of the corresponding thermal variance spectrum is shown in Figs. \ref{fig:3}(b) which is averaged with respect to the azimuthal angle in the horizontal plane. We confirm by these plots how the wave number that corresponds to the maximum of the spectrum drifts steadily towards smaller values in the course of the evolution. Such a peak is absent in the kinetic energy spectra for $k_z=0$ taken with respect to horizontal velocity components. The size of the horizontal temperature patterns during this aggregation process is quantified by the corresponding integral length scale \cite{Parodi2004} given by
\begin{equation}
\Lambda = 2 \pi\, \frac{ \int_0^{k_{h, \mathrm{max}}} (k_{h})^{-1} \bar{E}_{\Theta \Theta} (k_{h}, k_{z} = 0, t) \, dk_{h}} { \int_0^{k_{h, \mathrm{max}}} \bar{E}_{\Theta \Theta} (k_{h}, k_{z}= 0, t) \, dk_{h} }\,,
\label{intscale}
\end{equation}
with $\bar{E}_{\Theta\Theta}(k_{h},k_z=0,t)$ representing the azimuthally averaged thermal variance spectrum with respect to the horizontal wave number $k_{h}$ (see ref. \cite{Vieweg2021}). Note here that we decompose the wave number vector ${\bm k}=({\bm k}_{h}, k_z)$ with $\bm{k}_{h} = (k_{x}, k_{y})$ and $k_{h} = | \bm{k}_{h} |$. The wave number corresponding to this mean characteristic scale $\Lambda$ is included in panels (b)--(f) as solid vertical khaki lines. It is seen that this wave number moves to the left, indicating the continued aggregation of the temperature field patterns. This process is ceased once the domain size is reached, i.e., the temperature spectrum peaks at $k_{\mathrm{min}} = 2 \pi / \Gamma \approx 0.1$.

\begin{figure}[ht!]
\begin{center}
\includegraphics[scale=1.0]{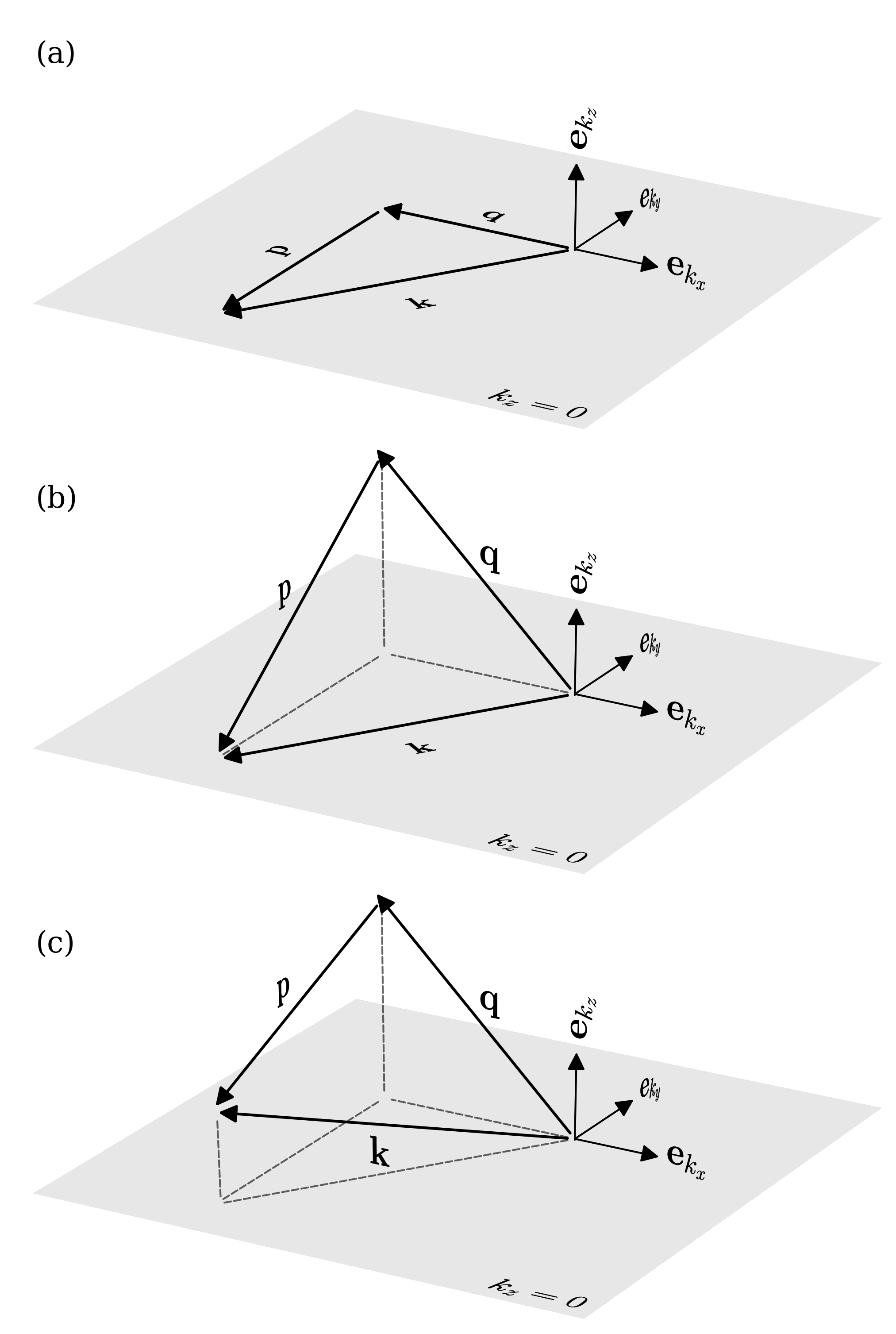}
\caption{Subsets of possible wave vector triads ${\bm k}={\bm p}+{\bf q}$ for the coupling of Fourier modes in spectral space. The reference wave vector is ${\bm k}$. (a) Wave vector triad in the horizontal plane (indicated in gray) for the coupling of vertically homogeneous Fourier modes with $k_z=p_z=q_z=0$. (b) Triad consisting of two three-dimensional wave vectors ${\bm p}$ and ${\bm q}$ that end up in the $k_{z} = 0$ plane -- hence, $q_{z} = - p_{z}$. (c) Fully three-dimensional wave vector triad. Only subsets shown in (a) and (b) are of particular interest for the inverse cascades reported here.}
\label{fig:4}
\end{center}
\end{figure}

In order to detect and judge the cascade mechanisms in the flow, the transfer term due to advective interactions among the Fourier modes, $\hat{\bm u}({\bm k},t)$ and $\hat{\Theta}({\bm k},t)$, has to be analysed. In Appendix A, we list the discrete Fourier transformation, derive the equations of motion in the spectral space, and provide the resulting balance equations for the kinetic energy and thermal variance spectra. The transfer terms result from the advection terms in the original equations of motion and couple any three Fourier modes with wave number vectors ${\bm k}$, ${\bm p}$, and ${\bm q}$ that can form a triad with ${\bm k}={\bm p}+{\bm q}$. The resulting spectral transfer terms for the turbulent kinetic energy and the thermal variance,  $A_u$ and $A_{\Theta}$, are given by (see Appendix A or refs. \cite{Verma2019,Plunian2020}) 
\begin{equation}
\label{eq:N_u_N_Theta}
A_{\Phi} ({\bm k}, t) = - \sum_{{\bm p}, {\bm q}} S_{\Phi} ({\bm k} | {\bm p} | {\bm q}, t)
\end{equation}
with $\Phi = u, \Theta$ and the rates of mode-to-mode energy transfer
\begin{align}
\label{eq:S_u_convolution}
S_{u} ({\bm k} | {\bm p} | {\bm q}, t) &= \imag\, \left\{ \left[ {\bm k} \cdot \hat{\bm u} ({\bm q},t) \right] \hat{\bm u} ({\bm p}, t) \cdot \hat{\bm u}^{\ast} ({\bm k}, t) \vphantom{\hat{\Theta}} \right\}\,, \\
\label{eq:S_Theta_convolution}
S_{\Theta} ({\bm k} | {\bm p} | {\bm q}, t) &= \imag\, \left\{ \left[ {\bm k} \cdot \hat{\bm u} ({\bm q},t) \right] \hat{\Theta} ({\bm p}, t) \thickspace \thickspace \hat{\Theta}^{\ast} ({\bm k}, t) \right\}\,. 
\end{align}
Here, \imag indicates the imaginary part, the asterisk stands for the complex conjugate, and $\hat{\bm u}({\bm k},t)=\hat{\bm u}^{\ast}(-{\bm k},t)$ as well as $\hat{\Theta}({\bm k},t)=\hat{\Theta}^{\ast}(-{\bm k},t)$. 

Spectral flux terms sum all such advective transfer interactions up to a wave number $k'$. The fact that the supergranule aggregation process is captured by the vertically averaged temperature field (see again Fig. \ref{fig:3}(a)), which in turn corresponds to wave vectors with $k_{z} = 0$, underlines the importance of two-dimensional modes. This suggests that the analysis of the cascade process should be disentangled based on the dimensionality of wave vector triads in the three-dimensional space, see Fig. \ref{fig:4} for grouping the triads. Two particularly important subsets of spectral fluxes can be derived: (1) the spectral thermal variance flux $\Pi^{\mathrm{h}}_{\Theta} (k_{h}^{\prime}, t)$ due to wave vector triads in the horizontal plane $(k_z = p_z = q_z = 0)$; (2) the spectral thermal variance flux $\Pi^{\mathrm{v}}_{\Theta} (k_{h}^{\prime}, t)$ into the vertically homogeneous Fourier modes ($k_{z} = 0$) that results from an interaction with fully three-dimensional wave vectors ($p_{z} = - q_{z} \neq 0$). These corresponding spectral flux terms are given by 
\begin{align} 
\label{eq:def_spectral_flux_Theta_h}
\Pi^{\mathrm{h}}_{\Theta} (k_{h}^{\prime}, t) &= - \sum_{0 \leq k_{h} \leq k_{h}^{\prime}} \sum_{{\bm p}, {\bm q}} \left. S_{\Theta} ({\bm k} | {\bm p} | {\bm q},t) \vphantom{\sum} \right|_{\begin{subarray}{l}k_z = 0, \\ p_z = q_z = 0\end{subarray}}\,,\\
\label{eq:def_spectral_flux_Theta_v}
\Pi^{\mathrm{v}}_{\Theta} (k_{h}^{\prime}, t) &= - \sum_{0 \leq k_{h} \leq k_{h}^{\prime}} \sum_{{\bm p}, {\bm q}} \left. S_{\Theta} ({\bm k} | {\bm p} | {\bm q},t) \vphantom{\sum} \right|_{\begin{subarray}{l}k_z = 0, \\ p_z = - q_z \neq 0\end{subarray}}\,.
\end{align}
Similar expressions follow for the spectral kinetic energy fluxes $\Pi^{\mathrm{h}}_{u} (k_{h}^{\prime}, t)$ and $\Pi^{\mathrm{v}}_{u} (k_{h}^{\prime}, t)$. We will focus our subsequent discussion on these 4 transfer terms as the supergranule formation is most prominent in the $k_{z} = 0$ plane. Note also that the full three-dimensional transfer terms $\Pi_u(k'_{\mathrm{max}},t)=\Pi_{\Theta}(k'_{\mathrm{max}},t)=0$, as the kinetic energy and thermal variance are exchanged only among the Fourier modes by these terms. This conservation does not necessarily exist for subsets of triads.

Figures \ref{fig:3}(c) display the spectral thermal variance flux $\Pi^{\mathrm{h}}_{\Theta} (k_{h}^{\prime}, t)$ due to purely horizontal mode interactions at different times. Note that the first four columns represent instantaneous transient states, whereas the fifth column represents data from the steady regime which is averaged over $10^{3}$ free-fall times centered around $t = 3,500$ -- see also Fig. \ref{fig:2}(a). During the transient aggregation, we observe an extended range of wave numbers for which $\Pi_{\Theta}^{\mathrm{h}}<0$. This proves the existence of an inverse cascade among the planar wave vectors causing the aggregation. As the wave number $k_{h}^{\prime}$ increases in each of the panels, the sign of $\Pi^{\mathrm{h}}_{\Theta}$ changes, thus indicating a direct cascade present at the smaller turbulence scales. Once the formation finishes (as the pair of convection cells reaches the domain size), the inverse cascade for the thermal variance extinguishes and $\Pi_{\Theta}^{\mathrm{h}}>0$ across the whole wave number range, see the rightmost panel of Figs. \ref{fig:3}(c).  

Figures \ref{fig:3}(d) show the spectral thermal variance flux $\Pi^{\mathrm{v}}_{\Theta} (k_{h}^{\prime}, t) $ which results from couplings with vertically non-homogeneous modes. It is seen that this quantity increases from small to large wave numbers; it corresponds to a direct cascade as $\Pi^{\mathrm{v}}_{\Theta}(k_{h}^{\prime})>0$ across all $k_{h}^{\prime}$ and exhibits a footprint of the time-dependent horizontal pattern size. This behavior is predominantly caused by interactions with $p_{z} = \pm \pi$ modes. Symmetric triads with higher-order ${\bm p}$ values remain subdominant compared to the ones with $p_{z} = \pm \pi$. 

\begin{figure}[ht!]
\begin{center}
\includegraphics[scale=1.0]{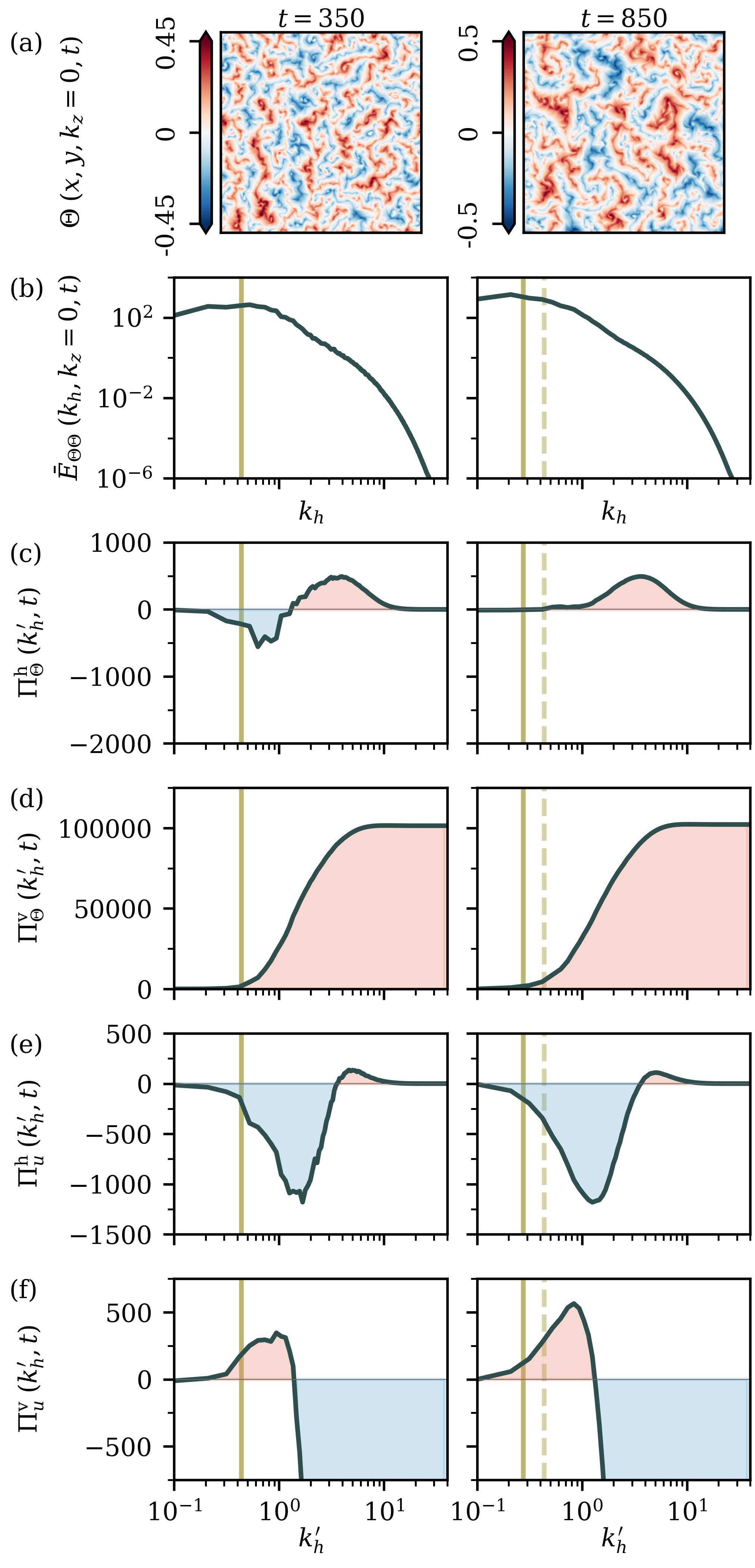}
\caption{Spectral transfer analysis for the weakly rotating convection case R1\_Ro5 at ${\rm Ra}=1.0\times 10^4$ and ${\rm Ro}=5$. The left column shows data for one instant time during the transient growth of the supergranule structure, whereas the right one provides time-averaged data (over $500 \tau_{f}$) from the statistically steady regime. For more information, we refer to the caption of Fig. \ref{fig:3}.}
\label{fig:5}
\end{center}
\end{figure}

Hence, it is the spectral flux $\Pi^{\mathrm{h}}_{\Theta}$ -- caused by purely horizontal mode interactions -- that explains an inverse cascade of thermal variance. The involved triads also comprise velocities from the $k_{z} = 0$ plane, so the above analysis is repeated for the turbulent kinetic energy.

In Figs. \ref{fig:3}(e), the results for $\Pi^{\mathrm{h}}_u (k_{h}^{\prime}, t)$ are shown. We identify a pronounced inverse cascade similar to the case of the thermal variance, but this cascade does not move over time to the largest scales. This inverse cascade exists for $k_{h}' \lesssim 3$ which corresponds to large-scale structures of horizontal size larger than $2H$. This is in contrast to the behavior of $\Pi^{\mathrm{v}}_u (k_{h}^{\prime}, t)$ as shown in Figs. \ref{fig:3}(f)  -- this spectral flux displays an inverse cascade for wave numbers $k_{h}^{\prime}\gtrsim 1$ which corresponds to spatial scales smaller than $6H$. It manifests all the transfer processes that lead to the establishment of the granules and the related larger-scale circulations. Interestingly, both of these inverse sub-cascades of the kinetic energy remain intact after the turbulent flow reaches the statistically stationary regime with respect to both, the turbulent kinetic energy and thermal variance. These two fluxes generate velocity structures with $k_{z} = 0$ which then enter $\Pi^{\mathrm{h}}_{\Theta}$.
The results remain the same when we switch to higher Rayleigh numbers in cases R2, R3 or R4. We provide details on R2 and R4 in Figures \ref{SI_fig1} and \ref{SI_fig2} in Appendix B. 

\section{Inverse cascade breakdown for weak rotation.}  
We now add (weak) rotation around the vertical axis to the dynamics of the convection layer.  Weak rotation implies that the dimensionless Rossby number remains well above unity, in our case $\Ro \gtrsim 2.5$, see e.g. \cite{Stevens2009}. This distinguishes the present cases from several investigations in the geostrophic regime of convection at ${\rm Ro}\ll 1$, which we will address in detail in the final discussion section. Figure \ref{fig:1} already demonstrated that the aggregation process to a supergranule is interrupted when the rotation is sufficiently strong. Thus, different characteristic scales of the filamented patterns follow for differently strong rotation rates when the global perspective is taken. Note that the corresponding characteristic horizontal scale of the pattern remains still significantly larger than the height $H$ of the layer, so still $\Lambda\gg 1$. 

Figure \ref{fig:5} displays the spectral transfer of run R1\_Ro5, the weakly rotating counterpart to run R1,  at two time instants. This simulation starts again with the perturbed, diffusive equilibrium state, but includes the Coriolis force term. Figs. \ref{fig:5}(a) demonstrate that additional weak rotation prevents the formation of one domain-sized large-scale flow structure. The inverse cascade of thermal variance in the subset of two-dimensional modes is at work as long as the convection patterns have to be rebuilt up to their new mean scale $\Lambda({\rm Ro})$. While this is still the case at the transient time $t=350$, the inverse cascade ceases eventually at $t=850$ as seen in Figs. \ref{fig:5}(c).  Once they have reached the maximal possible extension, the inverse cascade of the thermal variance vanishes with $\Pi_{\Theta}^{\mathrm{h}}(k_{h}',t)\ge 0$ across all horizontal limit wave numbers $k_{h}'$. This is in contrast to the cascades of the kinetic energy. The spectral transfer for the kinetic energy with respect to both subsets (1) and (2) -- as well as for the thermal variance with respect to the three-dimensional triads of subset (2) -- is not altered qualitatively from the non-rotating case.

\begin{figure}[t]
\begin{center}
\includegraphics[scale=1.0]{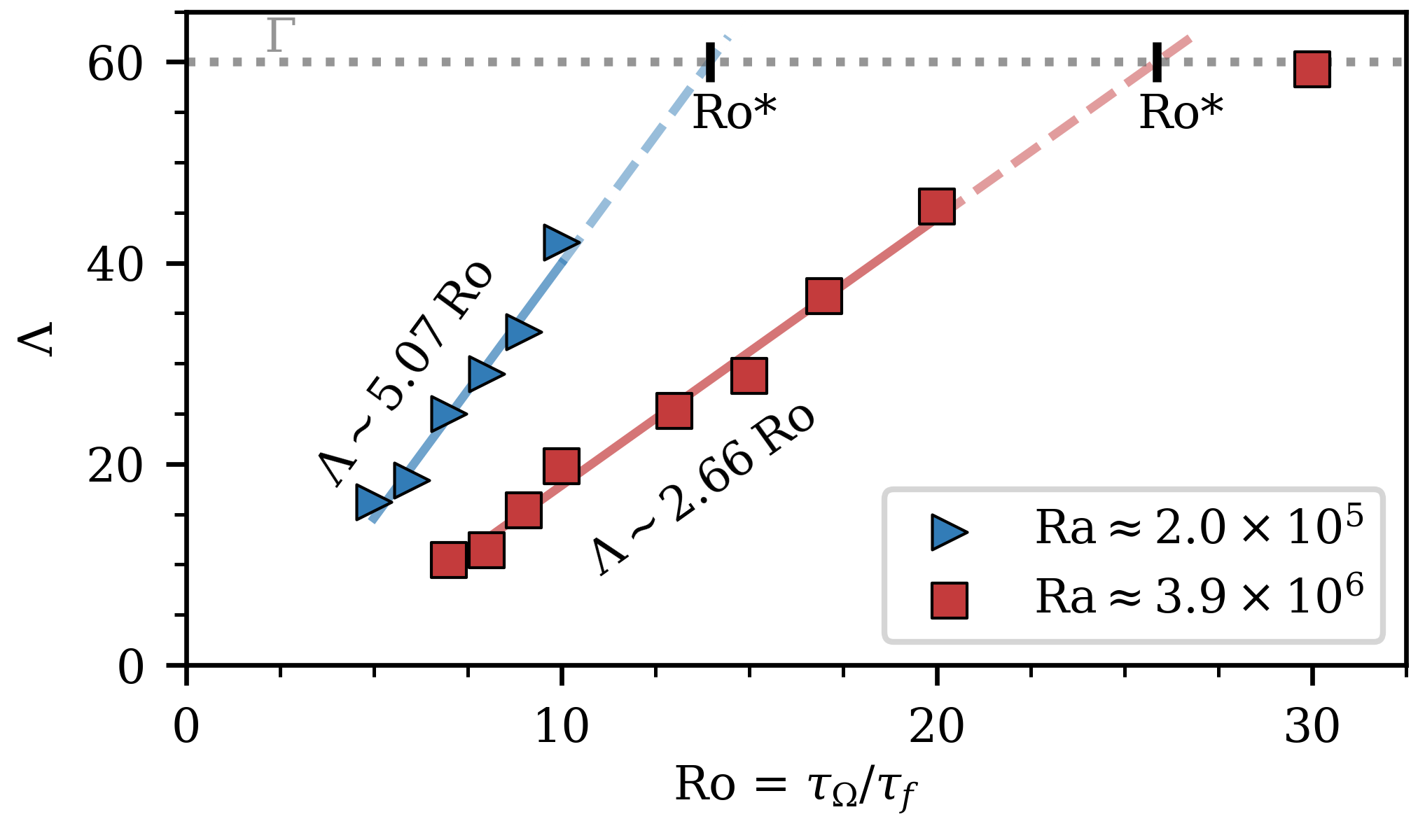}
\caption{Scaling of horizontal integral length scale for two series of simulations at varying Rossby and fixed Rayleigh number. Each marker represents the resulting characteristic horizontal scale $\Lambda$ of one simulation run at $\Gamma = 60$. The linear trend is highlighted by the solid lines in the background while the scaling coefficients are given for each series. Extrapolations up to domain size are provided as dashed lines. }
\label{fig:6}
\end{center}
\end{figure}

While Fig. \ref{fig:1} demonstrated that the final supergranule size depends on the strength of rotation, Figure \ref{fig:6} reveals that the characteristic horizontal pattern scale is in fact a \textit{linear} function of the Rossby number, 
\begin{equation}
\Lambda(\Ro) \approx a(\Ra) \, \Ro + b(\Ra) \;\; \mbox{for} \;\; 2.5\lesssim \Ro \lesssim \Ro^{\ast}\,.
\label{rot}
\end{equation} 
Here, $\Ro^{\ast}$ is a Rayleigh number-dependent Rossby number beyond which Coriolis forces will have no impact on the aggregation for the present case. Thus $\Lambda$ takes the value of the non-rotating case, $\Lambda=\Gamma$. Extrapolations yield values of $\Ro^{\ast}\approx 14$ for $\Ra=2.0\times 10^5$ and $\Ro^{\ast}\approx 26$ at $\Ra=3.9\times 10^6$, see figure \ref{fig:6}. This is furthermore confirmed by simulation run R3\_Ro30 at $\Ro=30$, for which the rotation is too weak to fragment the domain-sized supergranule into smaller filaments.

This linear scaling in \eqref{rot} can be understood by the following argument. The nonlinear convection term which is responsible for the cascade will be constrained by the Coriolis term in magnitude, i.e., $| \left( {\bm u}\cdot {\bm \nabla} \right) {\bm u}|\sim 2|{\bm \Omega}\times {\bm u}|$. For the granule of size $\Lambda$ this translates to $U_f^2/\Lambda\sim 2\Omega U_f$ or a $\Lambda$-based Rossby number of $\Ro_{\Lambda}\sim 1$. A re-translation to the original definition of the Rossby number in eq. \eqref{params} results in $\Lambda/H \sim \Ro$ which is observed in the figure. Note that $\Lambda$ is a dimensional quantity in this argumentation. From \cite{Vieweg2021} it is furthermore known that the supergranule aggregation in the non-rotating case proceeds slower with increasing $\Ra$ as the ratio of free-fall to diffusive time scales decreases. Hence, it can be expected that the slope $a$ in eq. \eqref{rot} depends on $\Ra$ as well. A replot of $\Lambda$ versus the Ekman number $\Ek=\Ro\sqrt{\Pr/\Ra}$ does not lead to a collapse of both curves. Further series at different Rayleigh numbers are required to draw a firm conclusion on the Rayleigh number dependence. 

Linear stability predicts for the non-rotating layer with Neumann conditions an onset of convection at a critical Rayleigh number $\Ra_c=120$ and a critical wave number $k_{c} = 0$ \cite{Hurle1967}. Interestingly, this wave number moves off zero once the rotation rate is sufficiently large \cite{Dowling1988,Takehiro2002}. For ${\rm Pr}\ge 0.67$ and a critical Taylor number $\Ta_c=180.15$, the authors in \cite{Dowling1988,Takehiro2002} find $k_c>0$. The critical Taylor number  translates via $\Ta=\Ra/(\Pr \Ro^2)$ to Rossby number thresholds of $\Ro\gtrsim 34$ and $\Ro\gtrsim 148$ for the series R2 and R3, respectively, beyond which the supergranules would continue to grow in an infinitely extended domain. The extrapolation of our data in Fig. \ref{fig:6} to these thresholds would lead to aspect ratios which cannot be studied with our computational capabilities. Thus we cannot investigate possible implications of this threshold for the turbulent regime. 

\begin{figure}[t]
\begin{center}
\includegraphics[scale=1.0]{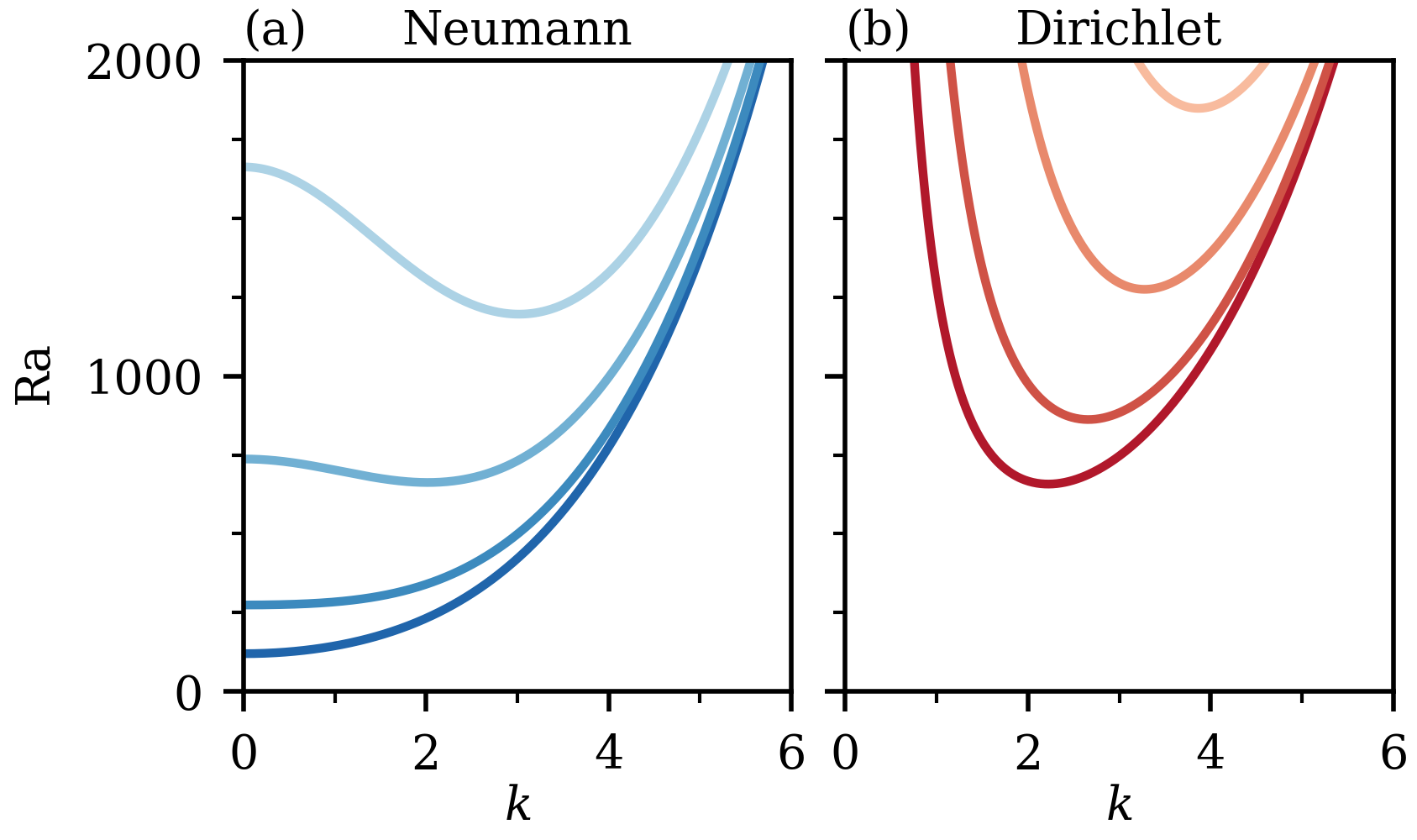}
\caption{Neutral stability curves of the linear stability analysis of the convection layer for (a) Neumann and (b) Dirichlet boundary conditions in combination with non-rotating and rotating cases. (a) Non-rotating Neumann case (dark blue) with ${\rm Ra}_c=120$ and $k_c=0$ and increasingly stronger rotating cases (in increasingly lighter blue) at Taylor numbers ${\rm Ta}=125, 500$ and 1250. (b) Non-rotating Dirichlet case (dark red) with ${\rm Ra}_c=27\pi^4/4$ and $k_c=\pi/\sqrt{2}$ and increasingly stronger rotating cases (in increasingly lighter red) at the same Taylor numbers as in (a). All cases are for free-slip boundary conditions of the velocity field. The Taylor number ${\rm Ta}$ is an alternative measure of the strength of rotation. The larger ${\rm Ta}$ the stronger the layer rotates and the larger $k_c$ and ${\rm Ra}_c$. These curves are obtained for an infinite layer.}
\label{fig:7}
\end{center}
\end{figure}

Figure \ref{fig:7} shows neutral stability curves for free-slip boundary conditions of the velocity field. As just mentioned in the last paragraph, panel (a) implies that a sufficiently large rotation rate $\Omega$ shifts the critical wave number to $k_c>0$ in the Neumann case. Increasing rotation rates (and thus increasing ${\rm Ta}$) stabilize the layer further which is visible by increasing values of $k_c$ and ${\rm Ra}_c$ for the global minima of the neutral stability curves. Consequently, the very large spatial scales become increasingly stabilized as well and will not be preferentially selected as the pattern scales. The corresponding Dirichlet case gives a qualitatively different picture for absent or very weak rotation rates. The critical wave number is always larger zero, $k_c\ge \pi/\sqrt{2}$. We can also see from both panels that for sufficiently large rotation rates the neutral stability curves become qualitatively equal for both thermal boundary conditions in a range around the global minimum. Recall that all data points in the ${\rm Ra}$--$k$ parameter plane which are above or enclosed by the neutral stability curve can be selected by the turbulent flow. 

Finally, as we have shown in Fig. \ref{fig:1}, the {\em local} structure of the convective turbulence below $\sim$ 2-3 $H$, including the boundary layers, does not strongly differ for weakly rotating and non-rotating cases. The same should hold when we compare Dirichlet and Neumann cases. Plume detachments from the plates by boundary layer instabilities remain present as the local heat transfer processes. This point is also supported by Nusselt and Reynolds numbers that remain comparable \cite{Johnston2009,Vieweg2021}. The substantial difference of the Neumann to the Dirichlet case is that additional degrees of freedom are excitable in the dynamical system which correspond to the supergranules, i.e., structures that will always tend to fill the whole domain. This starts at the onset of convection and proceeds into the fully turbulent regime to Rayleigh numbers that we could access, as shown by a leading Lyapunov vector analysis in \cite{Vieweg2021}. Even for stronger rotation, the neutral stability curves are supported all the way to $k=0$. However, the required critical Rayleigh numbers increase and thus these degrees of freedom will become dynamically less significant in comparison to those close to the global curve minimum, see e.g. the neutral stability curve for ${\rm Ta}=1250$ in Fig. \ref{fig:7}(a).

\section{Discussion and outlook} 
The first result of the present work was to prove the existence of inverse energy cascades in spectral space in a fully turbulent three-dimensional Rayleigh-B\'{e}nard flow without additional rotation. These cascades lead to the formation of a pair of large-scale salient convection cells -- termed here supergranules -- which are superposed by a network of smaller-scale flow structures -- granules -- which fill the whole layer. The supergranules are due to the increase of thermal variance most prominently visible in the temperature field, but exist also in the advecting velocity field. We find that the supergranules are mostly captured in the vertically homogeneous temperature mode for $k_{z} = 0$. The related modes of the velocity field are mostly found for $k_z>0$. 

These supergranules are observed only for constant heat flux boundary conditions of the temperature field at the top and bottom of the horizontally extended layer. A spectral transfer analysis of the Fourier modes for free-slip velocity and Neumann temperature boundary conditions offers a negative spectral flux $\Pi^{\mathrm{h}}$ at large scales for both the turbulent kinetic energy and the thermal variance. These inverse cascades are established by advective couplings of three Fourier modes from the subset of vertically homogeneous modes with only horizontal dependencies. Other triadic couplings, including non-local ones with three-dimensional wave vectors, are possible, but they do not directly contribute to the aggregation process on large scales. In this case, the energy transfer decreases significantly with increasing scale separation, see also \cite{Alexakis2005}. Furthermore, $\Pi_{u}^{\mathrm{v}}$ allows the growth of small-scale velocity structures only up to intermediate horizontal extensions. We also confirmed that the inverse cascades of both spectral kinetic energy fluxes, $\Pi_{u}^{\mathrm{h}}$ and $\Pi_{u}^{\mathrm{v}}$, are balanced by the remaining terms in eq. \eqref{eq:modal_kinetic_energy_abbreviated} (see Appendix A) in the final statistically stationary state.

Remarkably, all the reported inverse cascades are present even in a simple Rayleigh-B\'{e}nard flow. Additional physical processes that typically constrain the energy transfer in one specific space direction, such as rotation or strong magnetic fields, are absent. The inverse cascade in $\Pi_{\Theta}^{\mathrm{h}}$ for the temperature is at work as long as the aggregation proceeds. It ceases once this process is completed, i.e., once the supergranule fills the entire convection layer. Furthermore, we stress again that the injection of kinetic energy and thermal variance into the turbulent flow proceeds in a much more "uncontrolled" -- and thus natural -- way as in previous studies of two-dimensional or thin-layer turbulence  \cite{Smith1993,Frishman2017,Musacchio2019,vanKan2019}. The width of the thermal plumes, which detach permanently and spontaneously from the bottom and top boundaries, varies over a whole range and thus defines a locally varying energy or variance injection scale \cite{Scheel2014,Schumacher2015}. 

It is peculiar that such an aggregation process cannot be observed in case of Dirichlet boundary conditions. The crucial difference here is that this thermal boundary condition does not offer a $k_{z} = 0$ mode (as $\Theta$ must be expanded as sine function here; see more on this in Appendix A), and thus does not allow for interactions through $\Pi_{\Theta}^{\mathrm{h, v}}$. 

The present analysis of the spectral flux in triad subsets also extends earlier reported investigations at an aspect ratio $\Gamma=1$ in refs. \cite{Kumar2014,Verma2017}. For the fully three-dimensional spectral flux $\Pi_{u} (k', t)$ of the non-rotating cases we confirm a direct cascade for $k^{\prime}>\pi$, see also Fig. \ref{fig:4}(c). For wave numbers $k^{\prime}<\pi$, only triads that are illustrated in Figs. \ref{fig:4}(a,b) can contribute.

\begin{figure}[t]
\begin{center}
\includegraphics[scale=1.0]{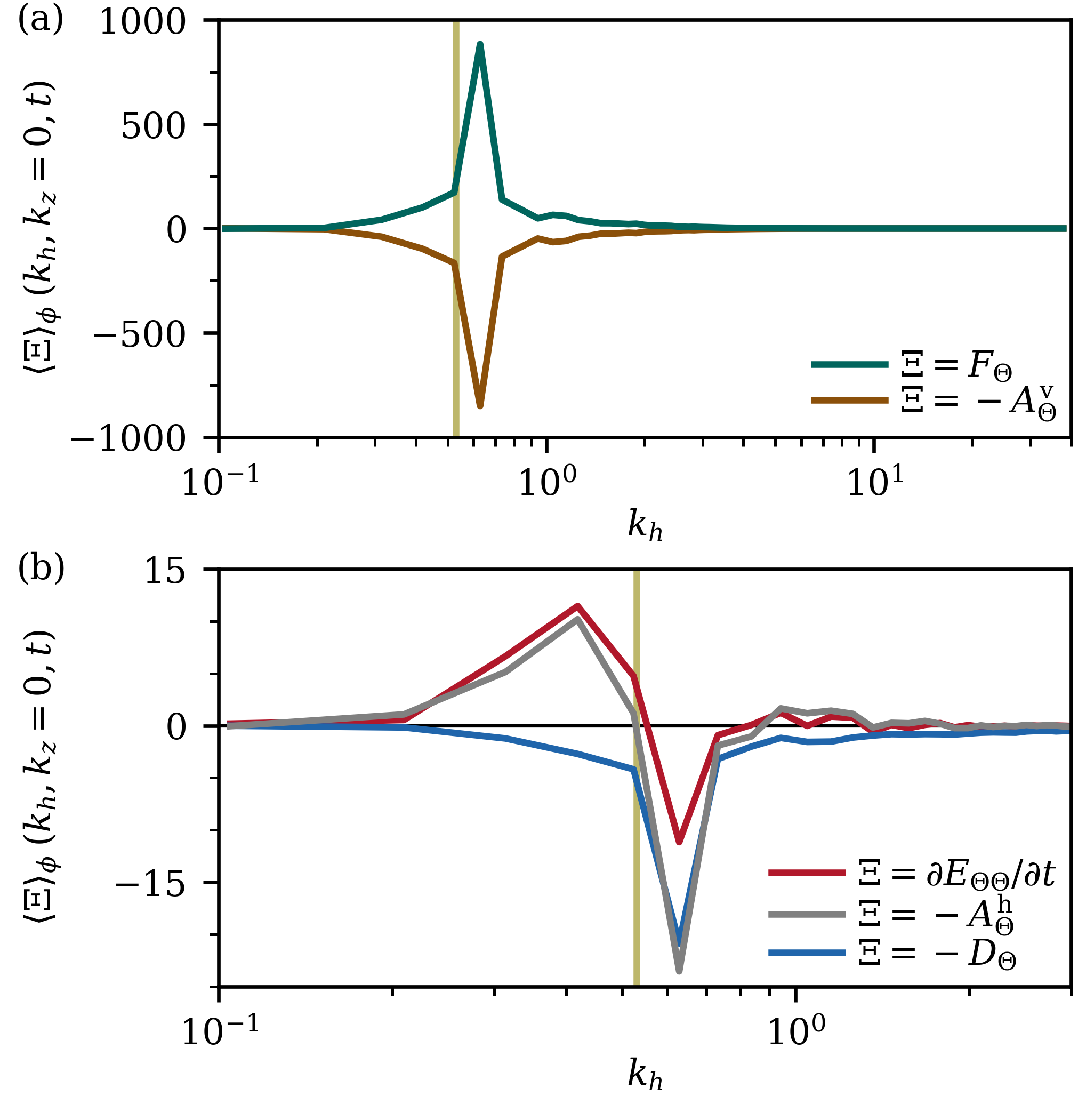}
\caption{Analysis of azimuthally averaged terms of the spectral balance of the thermal variance versus wave number, see  eq. \eqref{eq:modal_thermal_energy_abbreviated} in Appendix A. Data are for the simulation run Nfs1 at time $t = 350$ and $k_{z} = 0$, see also the first column of Fig. \ref{fig:3}. The solid vertical line quantifies the current supergranule size. The advection term is split similar to the spectral flux in eq. \eqref{eq:def_spectral_flux_Theta_h} and \eqref{eq:def_spectral_flux_Theta_v}.}
\label{fig:8}
\end{center}
\end{figure}

One might ask whether the growth of the kinetic energy and thermal variance for modes with $k_{z} = 0$ could be caused by the forcing terms in the spectral balance equations (see Appendix A). This is not the case for two reasons. Firstly, the kinetic energy forcing term $F_{u} (\bm{k}_{h}, k_{z} = 0, t) = 0$ for all $\bm{k}_{h}$ and thus there is no contribution to the time derivative of the spectrally resolved kinetic energy in eq. \eqref{eq:modal_kinetic_energy_abbreviated}. Secondly, we see in Fig. \ref{fig:8}(a) that $\langle F_{\Theta} \rangle_{\phi} \left( k_{h}, k_{z} = 0, t \right) \simeq \langle A_{\Theta}^{\mathrm{v}} \rangle_{\phi} \left( k_{h}, k_{z} = 0, t \right)$ at transient times where $\langle \cdot\rangle_{\phi}$ is an azimuthal average. In contrast, we find that $\langle A_{\Theta}^{\mathrm{h}} \rangle_{\phi} \left( k_{h}, k_{z} = 0, t \right) \simeq - \langle \partial E_{\Theta \Theta}/\partial t \rangle_{\phi} \left( k_{h}, k_{z} = 0, t \right)$, which underlines that it is the nonlinear mode coupling that accounts mostly for the temporal growth of the thermal variance spectrum, see Fig. \ref{fig:8}(b).

The second result of the present work was to demonstrate how the increase of the complexity of the convection flow by an inclusion of a Coriolis force term opens the door to control the size of the horizontal convection patterns -- the inverse cascade in $\Pi_{\Theta}^{\mathrm{h}}$ thus extinguishes eventually due to the effect of rotation. This additional mechanism is able to interrupt the aggregation process at a scale that is on the one hand well below the horizontal length of the domain, but on the other hand still much larger than the height of the layer. This demands direct numerical simulations to be run for an extraordinary large aspect ratio domain, which was chosen here to be $\Gamma=60$ for all runs. 

As mentioned throughout the text, the present study is related to the non-rotating or weakly rotating scenario and thus differs from a number of previous large-scale structure investigations performed in the rapidly rotating (or geostrophic) case for $\Ro\ll 1$ in domains with aspect ratio $\Gamma\sim 1$ (and partial inclusions of magnetic fields) \cite{Julien2012,Favier2014,Guervilly2014,Stellmach2014,Calkins2015,Kolhey2022} -- see also \cite{Aurnou2020} for a discussion of different rotation regimes. In particular, Julien et al. \cite{Julien2012} found that the formation of a large vortical structure is also caused by an inverse cascade, whereas Favier et al. \cite{Favier2014} showed that this appears for a finite Rayleigh number range only.  The geostrophic convection is thus different from the present one since we have observed the aggregation for all accessible Rayleigh numbers. Moreover, the main axis of the supergranules lies in the horizontal direction as they represent basically horizontally extended convection rolls. 

Our study stresses the fundamental importance of two-dimensional mode couplings within fully three-dimensional flows by proving the existence of inverse cascades in such complex flows that are subject to natural forcing. 
This significantly expands the set of known constellations that allow for inverse cascades towards even more natural scenarios. Our study thus contributes to a better understanding of structure formation in atmospheric and astrophysical convection processes where heat fluxes are typically prescribed at the top and bottom boundaries of the convection zone. An interesting point for the future work would be to extend the analysis to lower Prandtl numbers \cite{Vieweg2021a} as is the case for weakly rotating solar convection \cite{Vasil2021, Schumacher2020}. A further interesting point is the effect of stratification across the layer on the supergranule formation. This will require a non-Boussinesq convection model.

\section{Acknowledgements}
The work of PPV is supported by the Deutsche Forschungsgemeinschaft within the Priority Programme DFG-SPP 1881 on Turbulent Superstructures. The work of JDS is supported by a Mercator Fellowship of the DFG. The authors gratefully acknowledge the Gauss Centre for Supercomputing e.V. (www.gauss-centre.eu) for funding this project by providing computing time through the John von Neumann Institute for Computing (NIC) on the GCS Supercomputer JUWELS at J\"ulich Supercomputing Centre (JSC). We thank Laurent Gizon, Susanne Horn, Detlef Lohse and Moritz Linkmann for discussions.

\appendix
\section{Appendix A: Fourier expansion and spectral transfer}
In order to keep the work self-contained we list the basic relations for the spectral transfer analysis. A comprehensive discussion can be found for example in refs. \cite{Verma2017,Verma2019}.  The turbulence fields are spectrally expanded in accordance with periodic boundary conditions in the lateral directions $x$ and $y$ plus free-slip and constant heat flux boundary conditions in the vertical direction $z$, see eqns. \eqref{eq:free_slip_BC} and \eqref{eq:thermal_BC_Theta} in the main text. We denote the spatial vector by ${\bm x}=(\bm{x}_{h},z)$ and the wave number vector by ${\bm k}=(\bm{k}_{h},k_z)$ where the vectors with the subscript $h$ capture both horizontal components. Since $z\in [0,1]$ it follows that the discrete vertical wave numbers $k_z=n\pi$ with $n=0,1,2,\dots$. The Fourier expansions are thus given by
\begin{alignat}{4} 
\label{eq:spectral_expansion_ux}
u_{x} \left( \bm{x}, t \right) 	&= \sum_{\bm{k}_{h}} \sum_{k_{z}} \thickspace \hat{u}_{x} &\left( \bm{k}_{h}, k_{z}, t \right) \thickspace 	&e^{\imath \bm{k}_{h} \cdot \bm{x}_{h}} \thickspace &\cos \left( k_{z} z \right) , \\
\label{eq:spectral_expansion_uy}
u_{y} \left( \bm{x}, t \right) 	&= \sum_{\bm{k}_{h}} \sum_{k_{z}} \thickspace \hat{u}_{y} &\left( \bm{k}_{h}, k_{z}, t \right) \thickspace 	&e^{\imath \bm{k}_{h} \cdot \bm{x}_{h}} \thickspace &\cos \left( k_{z} z \right) , \\
\label{eq:spectral_expansion_uz}
u_{z} \left( \bm{x}, t \right) 	&= \sum_{\bm{k}_{h}} \sum_{k_{z}} \thickspace \hat{u}_{z} &\left( \bm{k}_{h}, k_{z}, t \right) \thickspace 	&e^{\imath \bm{k}_{h} \cdot \bm{x}_{h}} \thickspace &\imath \sin \left( k_{z} z \right) , \\
\label{eq:spectral_expansion_Theta}
\Theta \left( \bm{x}, t \right) &= \sum_{\bm{k}_{h}} \sum_{k_{z}} \thickspace \, \hat{\Theta} &\left( \bm{k}_{h}, k_{z}, t \right) \thickspace 	&e^{\imath \bm{k}_{h} \cdot \bm{x}_{h}} \thickspace &\cos \left( k_{z} z \right) .
\end{alignat}
Note that $\Theta$ would have to be expanded in sine functions with respect to the vertical coordinate $z$ for Dirichlet temperature boundary conditions. Thus Fourier modes with $k_z=0$ are zero in contrast to the present case. Also, the notation in \eqref{eq:spectral_expansion_ux}--\eqref{eq:spectral_expansion_Theta} is chosen such that it is compatible with a Fourier expansion in a triply periodic domain for fields that are even or odd with respect to $z$-coordinate.

Equations \eqref{eq:spectral_expansion_ux}--\eqref{eq:spectral_expansion_Theta} are substituted into the Boussinesq equations \eqref{eq:CE} --\eqref{eq:EE}. After a projection onto the basis functions of a particular wave vector, an infinite-dimensional system of coupled non-linear ordinary differential equations for the Fourier modes is obtained. For a wave vector ${\bm k}$, it is given by 
\begin{align}
\label{eq:CE_spectral}
\bm{k} \cdot \bm{\hat{u}} = 0 \qquad \qquad \qquad
\\
\begin{aligned}
\label{eq:NSE_spectral}
\frac{\partial \bm{\hat{u}}}{\partial t} 
+ \imath {\bm k} \cdot \thickspace \widehat{\bm{u} \bm{u}} 
+ \frac{1}{\Ro} \thickspace \bm{e}_{z} \times \bm{\hat{u}}
= \qquad \\ \qquad
- \imath \bm{k} \thickspace \hat{p} 
- \sqrt{\frac{\Pr}{\Ra}} \thickspace k^{2} \bm{\hat{u}} 
+ \hat{\Theta}_{\textrm{SE}} \thickspace \bm{e}_{z}
\end{aligned}
\\
\label{eq:EE_spectral}
\frac{\partial \hat{\Theta}}{\partial t} 
+ \imath {\bm k} \cdot \thickspace \widehat{\bm{u} \Theta} 
= 
- \frac{1}{\sqrt{\Ra \Pr}} \thickspace k^{2} \hat{\Theta} 
+ \hat{u}_{z, \textrm{CE}}
\end{align}
The widehat symbols stand for convolution sums which are not written out in order to keep a compact notation. The Fourier modes $\hat{u}_{z, \textrm{CE}}$ and $\hat{\Theta}_{\textrm{SE}}$ represent projections of the corresponding quantity onto cosine (CE) or sine (SE) basis functions that become necessary because of the constant flux boundary conditions and the spectral expansions. The equations for the kinetic energy or thermal variance of a Fourier mode are then given by
\begin{align}
\label{eq:modal_kinetic_energy_abbreviated}
\frac{\partial E_{uu}({\bm k}, t)}{\partial t} = \frac{1}{2} \frac{\partial}{\partial t}|\bm{\hat{u}} ({\bm k},t)|^2 &= - A_{u} - D_{u} + F_{u} ,\\
\label{eq:modal_thermal_energy_abbreviated}
\frac{\partial E_{\Theta \Theta} ({\bm k}, t)}{\partial t} = \frac{1}{2} \frac{\partial}{\partial t}|\hat{\Theta}({\bm k},t)|^2&= - A_{\Theta} - D_{\Theta} + F_{\Theta} 
\end{align}
with 
\begin{align}
\label{eq:abbreviation_A_u_long}
A_{u} \left( \bm{k}, t \right)		&= \real \left[ (\imath \bm{k}\cdot \widehat{\bm{u} \bm{u}})  \cdot \bm{\hat{u}^*} \right] ,\\
\label{eq:abbreviation_A_Theta_long}
A_{\Theta} \left( \bm{k}, t \right)	&= \real \left[ (\imath\bm{k}\cdot \widehat{\bm{u} \Theta}) \,\hat{\Theta}^* \right] ,\\
D_{u} \left( \bm{k}, t \right)		&= \sqrt{\frac{\Pr}{\Ra}} \thickspace k^{2} \thickspace |\bm{\hat{u}} \left( \bm{k}, t \right)|^2 ,\\
D_{\Theta} \left( \bm{k}, t \right)	&= \frac{1}{\sqrt{\Ra \Pr}} \thickspace k^{2} \thickspace |\hat{\Theta}\left( \bm{k}, t \right)|^2 ,\\
F_{u} ({\bm k}, t) &= \real \left[ \hat{\Theta}_{\textrm{SE}} \thickspace \hat{u}_{z}^* \right] ,\\
F_{\Theta} ({\bm k}, t) &= \real \left[ \hat{u}_{z, \textrm{CE}} \thickspace \hat{\Theta}^* \right] .
\end{align}
The advective transfer terms \eqref{eq:abbreviation_A_u_long} and \eqref{eq:abbreviation_A_Theta_long} can finally be expressed as a convolution such that 
\begin{align}
\label{eq:convolution_A_Theta_imag}
A_{\Theta} ({\bm k}, t) &= - \imag \left\{ \sum_{\bm{p}, \bm{q}} \left[ {\bm k}\cdot\hat{\bm u}({\bm q},t) \right] \thickspace \hat{\Theta} ({\bm p}, t) \thickspace \hat{\Theta}^* ({\bm k}, t) \right\} ,\\
\label{eq:convolution_A_u_imag}
A_{u} ({\bm k}, t) &= - \imag \left\{ \sum_{\bm{p}, \bm{q}} \left[ {\bm k}\cdot\hat{\bm u}({\bm q},t) \right] \thickspace \hat{\bm{u}} ({\bm p}, t) \cdot \hat{\bm{u}}^* ({\bm k}, t) \right\} 
\end{align}
with $\bm{q} = \bm{k} - \bm{p}$. These quantities are at the focus of our study. See also eqns. \eqref{eq:N_u_N_Theta} -- \eqref{eq:S_Theta_convolution} in the main text.

\section{Appendix B: Inverse cascade at higher Rayleigh numbers}
\begin{figure*}[p]
\centering
\includegraphics[scale=1.0]{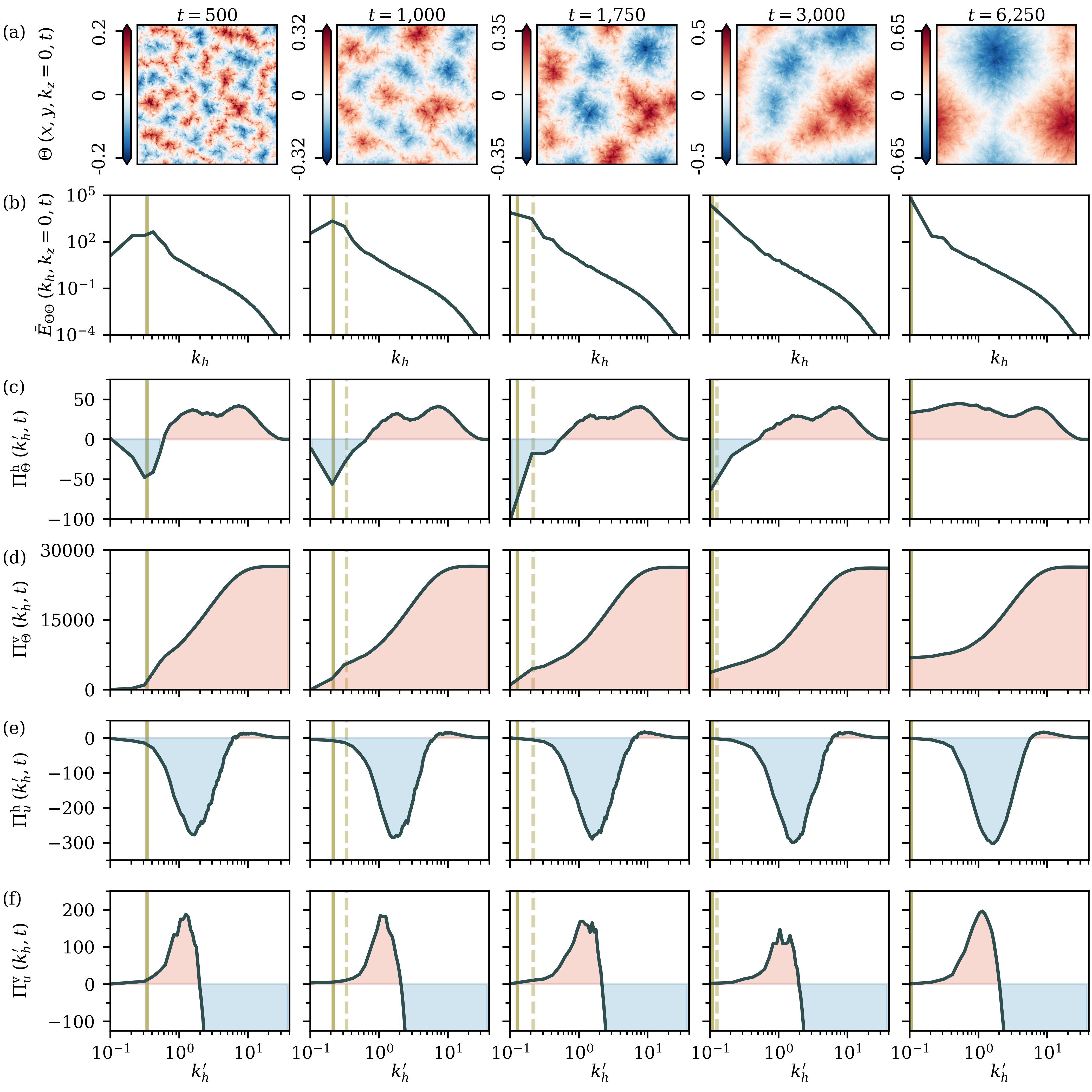}
\caption{Spectral transfer analysis for simulation R2 at a Rayleigh number $\Ra = 2.0 \times 10^{5}$ without rotation. 
(a) Instantaneous vertically averaged temperature field at different times. 
(b) Corresponding azimuthally averaged thermal variance spectrum. 
The solid vertical lines in these and all subsequent panels indicate the wave number that corresponds to the typical pattern size at the given time, whereas the dashed lines replot this quantity from the previous panel to highlight the growth of this length scale over time. 
(c) Planar spectral thermal variance flux $\Pi^{\mathrm{h}}_{\Theta} (k_{h}^{\prime}, t)$. 
(d) Three-dimensional spectral thermal variance flux $\Pi^{\mathrm{v}}_{\Theta} (k_{h}^{\prime}, t)$. 
(e) Planar spectral kinetic energy flux $\Pi^{\mathrm{h}}_u (k_{h}^{\prime}, t)$. 
(f) Three-dimensional spectral kinetic energy flux $\Pi^{\mathrm{v}}_{u} (k_{h}^{\prime}, t)$. 
The data in the last column are obtained when the supergranule growth reached domain-size and the turbulent flow becomes statistically stationary -- the rightmost panels of (b)--(f) are thus time-averaged over $500  \tau_{f}$ centered around the given time. In panels (b)--(f), blue (red) filled areas indicate an inverse (direct) cascade.}
\label{SI_fig1}
\end{figure*}

\begin{figure*}[p]
\centering
\includegraphics[scale=1.0]{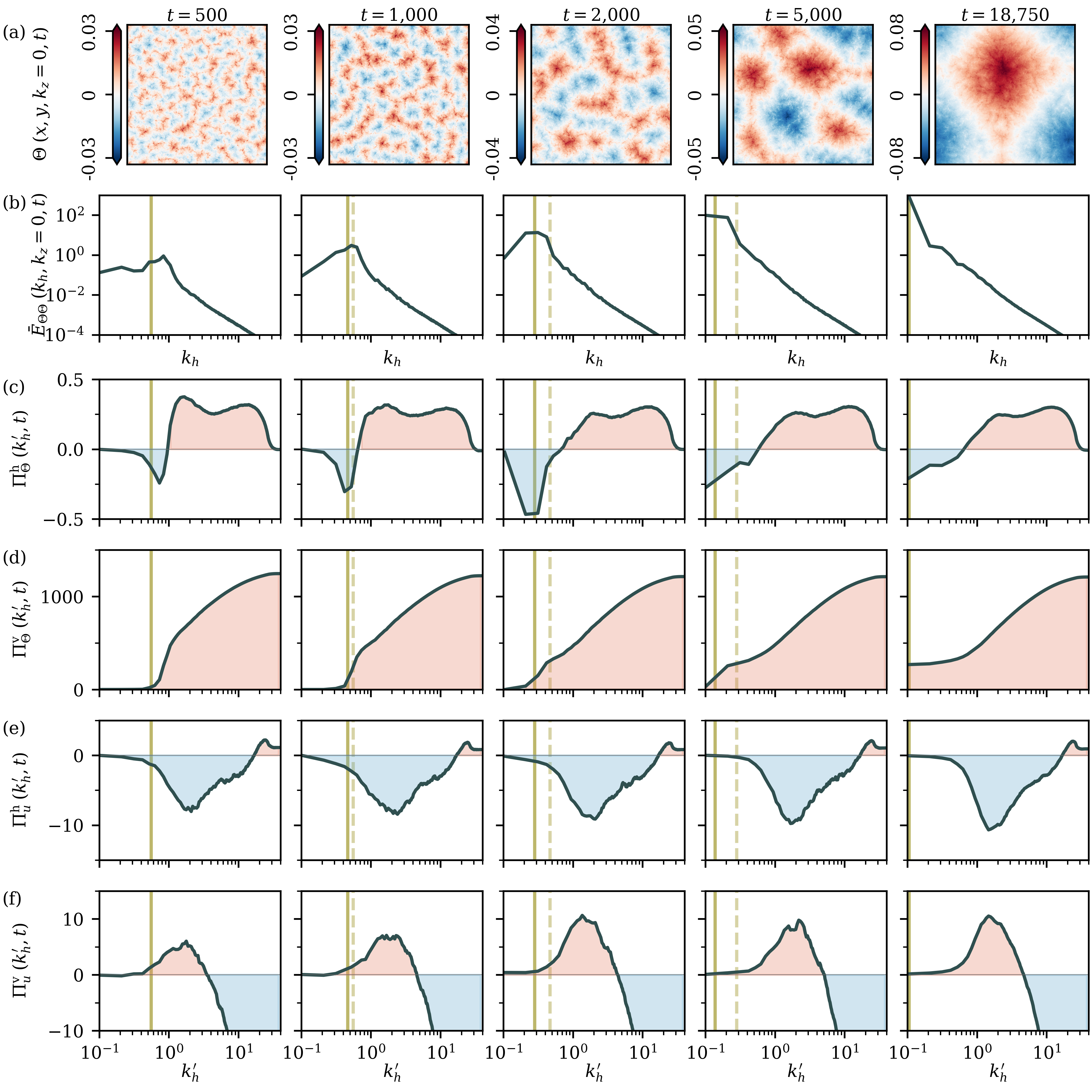}
\caption{Spectral transfer analysis for simulation R4 at a Rayleigh number $\Ra = 7.7 \times 10^{7}$ without rotation. 
(a) Instantaneous vertically averaged temperature field at different times. 
(b) Corresponding azimuthally averaged thermal variance spectrum. 
The solid vertical lines in these and all subsequent panels indicate the wave number that corresponds to the typical pattern size at the given time, whereas the dashed lines replot this quantity from the previous panel to highlight the growth of this length scale over time. 
(c) Planar spectral thermal variance flux $\Pi^{\mathrm{h}}_{\Theta} (k_{h}^{\prime}, t)$. 
(d) Three-dimensional spectral thermal variance flux $\Pi^{\mathrm{v}}_{\Theta} (k_{h}^{\prime}, t)$. 
(e) Planar spectral kinetic energy flux $\Pi^{\mathrm{h}}_u (k_{h}^{\prime}, t)$. 
(f) Three-dimensional spectral kinetic energy flux $\Pi^{\mathrm{v}}_{u} (k_{h}^{\prime}, t)$. 
The data in the last column are obtained when the supergranule growth reached domain-size and the turbulent flow becomes statistically stationary -- the rightmost panels of (b)--(f) are thus time-averaged over $500  \tau_{f}$ centered around the given time. In panels (b)--(f), blue (red) filled areas indicate an inverse (direct) cascade.}
\label{SI_fig2}
\end{figure*}

Figure \ref{SI_fig1} repeats the spectral transfer analysis for simulation run R2 from the main text which is located in the fully turbulent regime. The dynamical behavior is qualitatively the same as the case R1 from the main text. The inverse cascade in $\Pi_{\Theta}^{\mathrm{h}}$ ceases once the formation of the supergranule is completed. For the run R4, which is at the highest Rayleigh number and shown in Figure \ref{SI_fig2}, we observe again an inverse cascade. This simulation was run for a total time interval of  19,000 free-fall time units. This time lag was still not enough to observe the stop of the inverse cascade of the thermal variance, as can be seen in this figure.



\end{document}